\documentclass[aps,preprint,12pt]{revtex4}
\usepackage{amssymb}
\usepackage{amsmath}
\usepackage{graphicx}
\usepackage{epsfig,amsmath}

\begin{document}

\title{Kinetics of a network of vortex Loops in He II and a theory of
superfluid turbulence}
\author{Sergey K. Nemirovskii\thanks{%
email address: nemir@itp.nsc.ru}}
\affiliation{Institute of Thermophysics, Lavrentyev ave, 1, 630090, Novosibirsk, Russia,\\
Novosibirsk State University, Novosibirsk Russia}
\date{\today}

\begin{abstract}
A theory is developed to describe the superfluid turbulence on the base of
kinetics of the merging and splitting vortex loops. Because of very frequent
reconnections the vortex loops (as a whole) do not live long enough to
perform any essential evolution due to the deterministic motion. On the
contrary, they rapidly merge and split, and these random recombination
processes prevail over other slower dynamic processes. To develop
quantitative description we take the vortex loops to have a Brownian
structure with the only degree of freedom, which is the length $l$ of the
loop. We perform investigation on the base of the Boltzmann type
\textquotedblright kinetic equation\textquotedblright\ for the distribution
function $n(l)$ of number of loops with length $l$. This equation describes
a slow change of the density of loops (in space of their lengths $l$) due to
the deterministic equation of motion and due to fast random change because
of the frequent reconnections. By use of the special ansatz in the
\textquotedblright collision\textquotedblright\ integral we have found the
exact power-like solution $n(l)\varpropto l^{-5/2}$ to \textquotedblright
kinetic equation\textquotedblright\ in the stationary case. This solution is
not (thermodynamically) equilibrium, but on the contrary, it describes the
state with two mutual fluxes of the length (or energy) in space of sizes of
the vortex loops. The term \textquotedblright flux\textquotedblright\ means
just redistribution of length (or energy) among the loops of different sizes
due to reconnections. Analyzing this solution we drew several results on the
structure and dynamics of the vortex tangle in the turbulent \ superfluid
helium. In particular, we obtained that the mean radius of the curvature is
of the order of interline space. We also evaluated the full rate of the
reconnection events. Assuming, further, that the processes of random
collisions are the fastest ones, we studied the evolution of full length of
vortex loops per unit volume-the so-called vortex line density $\mathcal{L}%
(t)$. It is shown this evolution to obey the famous Vinen equation. The
properties of the Vinen equation from the point of view of the developed
approach had been discussed. Thus, depending on the temperature (and
independently on velocity) vortices either develop into the highly chaotic
turbulent state (low temperature), or degenerate into few smooth lines (high
temperature). This observation can be an alternative explanation for the
phenomenon discovered in Helsinki group (Nature 424, 1022--1025 (2003)).%
\newline
PACS-numbers: 67.25.dk, 47.37.+q, 05.20.-y

\end{abstract}

\maketitle

\section{INTRODUCTION AND SCIENTIFIC BACKGROUND}

 This paper is the cumulative exposition of a series of preliminary
results reported on various scientific meetings and partly published in \cite%
{NemirPRL06} and in \cite{NemirLTP06}. It concerns an important role of the
fusion and breakdown processes of vortex loops in the whole dynamics of a
network of vortex filaments in superfluid helium. A network of
one-dimensional singularities appears in various physical systems. As
examples we would point out vortices in quantum fluids both in turbulent
regimes (see, e.g., book \cite{Donbook} and papers \cite{Feynman},\cite%
{Vinen},\cite{NF}) and in a thermodynamically equilibrium\ state (\cite%
{Zurek96}, \cite{Nemir05}). Other examples are the flux tubes in
superconductors \cite{Kleinert}, dislocations in solids \cite{Nabarro67},
global cosmic strings \cite{Copeland98},\cite{Steer99} and polymer chains
\cite{Wiegel73}. A network of one-dimensional singularities greatly affects
many properties of the system where they appear, such as phase transition,
thermodynamic and flow properties, structure formation, etc. Therefore, the
study of the evolution of these networks is the actual problem.

The evolution of a network of chaotic sets of lines consists of two main
ingredients. The first is the motion of the elements of lines, due to
equations of the motion, different for each of the cases listed above. For
instance, the elements of vortex filaments in superfluid $^{4}$He move
obeying the Biot-Savart law supplemented by the friction force and the
external flow/counterflow if any. Cosmic strings move with the speed close
to the speed of light $c$ up to some geometrical factor\cite{Copeland98}.
Velocities of molecules in a polymer chain are determined either by thermal
fluctuations of background or by the velocity fluctuations of the
surrounding solvent.We will call this ingredient of the evolution as a
deterministic motion of lines.

Beside the motion of the each individual loops there is another very
important constituent of the whole dynamics, common for all systems, which
relates to the collision of loops, or intersection of elements of vortex
lines. During intersection of lines the very complicated process, related to
arrangement of the vortex core takes place \cite{Koplik}. However this
process is relatively short, therefore it is usually accepted that the
filaments instantly reconnect whenever they intersect each other. \
Reconnections of lines result in random fusion and breakdown (recombination)
of the loops \cite{uniform}. The processes of recombination are
schematically depicted in Fig. 1.
\begin{figure}[tbp]
\includegraphics[width=12cm]{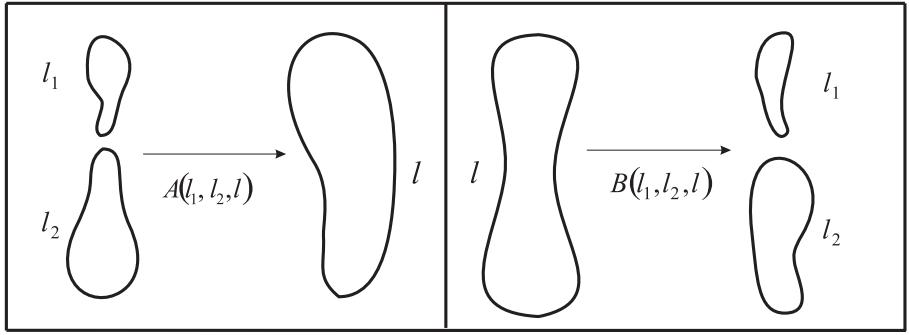}
\caption{Schematic sketch of fusion and breaking-down of vortex loops. Rates
of these processes characterized by the rate coefficients (number of events
per unit time and unit volume) $A(l_{1},l_{2},l)$ and $B(l,l_{1},l_{2})$.}
\end{figure}
On the left picture we depicted the process of fusion of two loops with
lengths $l_{1}$and $l_{2}$ and forming the loop with length $\ l=l_{1}+l_{2}$%
. On the right picture we depicted the self-intersection and break down of
loop of the length $l$ into two daughter loops with lengths $\ l_{1}$ and $%
l_{2}.$ The rates of these processes are characterized by the rate
coefficients $A(l_{1},l_{2},l)$ and $B(l,l_{1},l_{2})$ correspondingly.%
\newline
It is widely appreciated that the "recombination" processes greatly
influence both the structure and dynamics of the vortex tangle. For
instance, Feynman in his pioneering paper \cite{Feynman} devoted to
superfluid turbulence proposed that the vortex tangle decays due to the
cascade-like process of consequent breaking down of vortex loops, and
degenerating them into thermal excitations. This scenario is Schematically
depicted in Fig. 2. The Feynman's idea was confirmed in various numerical
calculations, where the procedure of artificial elimination of small loops
had been used \cite{Schwarz88}-\cite{Tsubota00}.\newline
\begin{figure}[tbp]
\includegraphics[width=12cm]{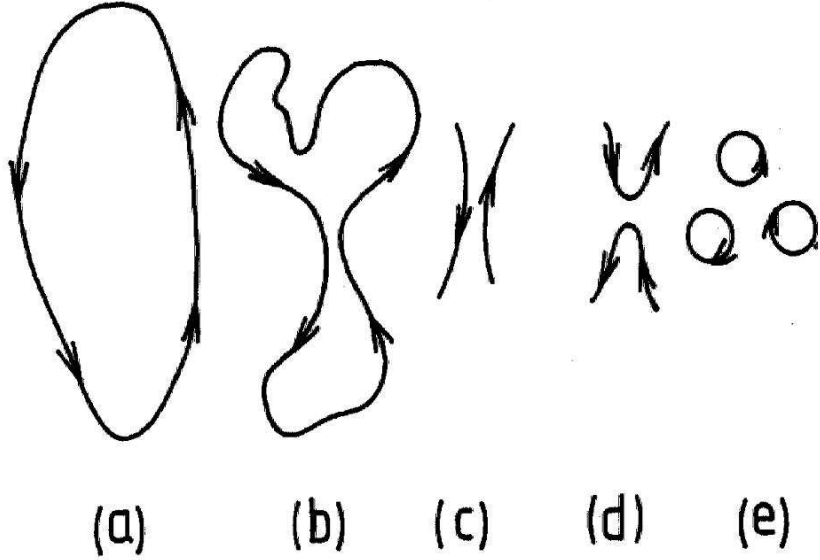}
\caption{Schematic sketch of cascade-like breakdown of vortex loops.
(Feynman, 1955, Fig.\ 10). (a) initial stage, (b) and (c) stages of
collapse, (d) reconnection stage, (e) stage of degeneration of vortex rings
into thermal excitations.}
\end{figure}
To clarify the role of recombination, let us perform the following numerical
estimation. The full rate of reconnection $\dot{N}_{rec}$ \ (per unit
volume) as a function of the vortex line density $\mathcal{L}$ had been
obtained in papers \cite{NemirPRL06},\cite{Barenghi2004} (see also
Subsection IV.3)
\begin{equation}
\dot{N}_{rec}=C_{rec}\kappa \mathcal{L}^{5/2},  \label{recon_rate}
\end{equation}%
where $C_{rec}$ is the constant of order of unity and $\kappa \approx
10^{-3} $ cm$^{2}$/s is the quantum of circulation. Let us take, for
instance, some typical experiments with superfluid turbulence, with the
counterflowing velocity of order of $1$ cm/s and with experimental volume of
order of $1$ cm. Under these conditions the typical value of the vortex line
density $\mathcal{L}$ is about $\mathcal{L}\approx 10^{4}\ 1/cm^{2}$.
Interline space $\mathcal{L}^{-1/2}$ is of order $10^{-2}\ cm$, this
quantity coincides with the mean radius of curvature. Then the full rate of
reconnection $\dot{N}_{rec}$ is of the order of $10^{7}$ collisions per
second (per unit volume).  Dividing it by $\mathcal{L}\approx
10^{4}\ 1/cm^{2}$ we obtain that the rate of reconnection per unit length of
the vortex filaments is $C_{rec}\kappa \mathcal{L}^{3/2}$ and is of the
order of $10^{3}\ 1/cm\ s$. Let us take \ a loop of length of ten of
interline space, $l\sim 10^{-1}\ cm$ \cite{loop_1cm}$.$ This loop undergoes
(on average) $10^{2}$ reconnections. per one second, or in other words it
exists (on average) $10^{-2}$ seconds without reconnection (as a whole). $%
\allowbreak $ On the other hand, the own vortex filament dynamics (Kelvin
waves dynamics) is a much slower process. For instances, if we take again a
loop with length $10^{-1}\ cm$, then any signal on the loop (for instance,
degradation of singularity appeared due to the reconnection event) takes
time about $l^{2}\ /\kappa \approx 10$ seconds. Thus, the characteristic
time of the Kelvin waves dynamics exceeds time of existence of the loop by $%
10^{3}$ times (!!!). If one takes a smaller loop the situation will be about
the same (with other quantitative estimations). In fact up to the smallest
loops of the size of interline space, the time of "life" without
reconnection is shorter than time of Kelvin wave running around the loop.
Only for the scale of the order of interline space $\mathcal{L}^{-1/2}$
these times are of the same order. But this means that loops (as a whole) do
not live long enough to perform any essential evolution due to the
deterministic motion. \emph{On the contrary, they frequently merge and
split, therefore these recombination processes are the fastest and \ the
basic approach to study of superfluid turbulence should be grounded on
consideration of a set of randomly merging and splitting loops.}{\Huge \ }

It is necessary to do the following two remarks. First, considering
superfluid turbulence as evolution of \ a network of vortex loops we
restrict ourselves to the scales of the least loops, which likely coincide
with interline space. We are not interested here in what is happening for
smaller scales. It is a separate topic connected to evolution of the bending
vibrations of vortex lines or the so-called Kelvin waves\cite{Nemirovskii93},%
\cite{Svistunov95}. This question is of a great interest from point of the
vortex tangle decay. Second, we do not consider here the case when the
vortex lines are strongly polarized so that the coarse-grained motion
induced by the bundles of filaments imitates classical (Kolmogorov)
turbulence (See \cite{Vinen00},\cite{Volovik03},\cite{Skrbek04}). We
consider here the case of the so-called Vinen turbulence, when the vortex
loops are highly disordered with zero mean vorticity.\newline
In spite of the recognized importance of the fusion and breakdown processes
for the evolution of a network of loops, the numerical results remain the
main source of information about this process. The obvious lack of
theoretical investigations interferes with the deep insight in the nature of
this phenomena (this situation had been recently discussed in \cite%
{Barenghi2004}). The scarcity of analytic investigations related to the
incredible complexity of the problem. Indeed, we have to deal with a set of
objects (vortex loops), which do not have a fixed number of elements, they
can be born and die. Thus, some analog of the secondary quantization method
is required with the difference that the objects (vortex loops) themselves
possess an infinite number of degrees of freedom with very involved
dynamics. Clearly, this problem can be hardly resolved in the nearest
future. Some approach crucially reducing the number of degrees of freedom is
required.\newline
There are various ways to overcome this problem. For instance, one approach
elaborated in context of lambda-transition (\cite{Williams}) is to think of
the vortex loops as a set of rings of different sizes and to take their
radius as the only degree of freedom. Another approach elaborated in the
context of \ cosmic strings (see \cite{Copeland98} and references therein)
is to imagine the vortex loops as having the Brownian or the random walk
structure. This can be motivated by the following consideration. Because of
the huge number of random collisions (see (\ref{recon_rate})) the structure
of any loop is determined by numerous previous reconnections. Therefore, any
loop consists of small parts, which "remember" previous collisions. This is
depicted schematically \ in Fig. 3.
\begin{figure}[tbp]
\includegraphics[width=12cm]{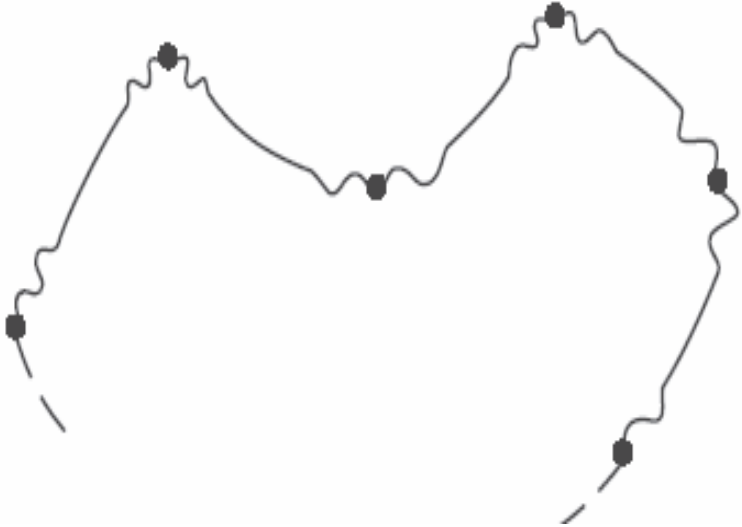}
\caption{Picture illustrating a Brownian nature of vortex loops. The
structure of any loop is determined by numerous previous reconnections
(black circles). Therefore, any loop consists of small parts which
"remember" previous collision. These parts are uncorrelated since
deterministic Kelvin wave signals (wavy parts near black circles) do not
have a time to propagate far enough. Thus, the loop has a structure of
random walk (like polymer chain).}
\end{figure}
Points indicate the sites where the previous reconnections occurred. Waves
on the curve are just Kelvin waves propagating from these sites. Fragments
of the line between the points of previous reconnections are uncorrelated
since deterministic Kelvin wave signals initiated by collisions of filaments
do not have time to propagate far enough. Therefore the loop has a structure
of the random walk (like a polymer chain).

The main mathematical tool to describe the random walk is the Wiener
distribution. We will use it in a form of the so-called generalized Wiener
distribution (See \cite{Nemirovskii_97_1} and Appendix A of the present
paper), which allows to take into account the finite curvature. Here we
consider the case of isotropic loops, omitting a possible variant of the
polarized loops having their own impulse. In this case the average loop can
be imagined as the consisting of many arches with the mean radius of
curvature equal $\xi _{0}$, randomly (but smoothly) connected to each other.
Quantity $\xi _{0}$ is the important parameter of the approach. It plays a
role of the "elementary step" in the theory of polymer. It is low cut-of of
the approach developed, the theory does not describe scales smaller then $%
\xi _{0}.$ Having quantity $\xi _{0}$ the only degree of freedom of the
random walk is the length of loop $l$. Resuming the said above we consider
the vortex tangle as a collection of vortex loops having various lengths $l$%
, and our goal now is to find distribution function $n(l)$ of the number of
loops in space of their lengths. Knowing quantity $n(l,t)$ and statistics of
each personal loop we are able to evaluate various properties of the real
vortex tangle

The paper is organized as follows. In the next, second Section we derive the
"kinetic" Boltzmann type equation for distribution function $n(l)$ and
specify the coefficients, entering it. In Section III, we obtain the
power-like stationary solution to the kinetic equation and describe its
properties. Section IV is devoted to the structure and dynamics of the real
vortex tangle in turbulent He II obtained on the base of the developed
approach. In particulary, we obtained analytically the equation for
evolution of the vortex line density (Vinen equation) and discuss its
properties. Two topics are relegated to appendices: the Gaussian model of
vortex loops (Appendix A) and detailed calculation of coefficients of the
kinetic equation (Appendix B).

\section{THE RATE EQUATION}

\subsection{Recombination of loops}

In introduction we exposed arguments that vortex loops composing the vortex
tangle have a random walk structure, which can be described with use of the
generalized Wiener distribution. We take parameters of this distribution not
to be changed while recombination, so the only degree of freedom of the loop
is its length $l$. This point of view coincides with conception elaborated
in paper \cite{Copeland98}, where similar problem had been studied in
context of cosmic strings. Following this work we introduce distribution
function $n(l,t)$, the density of loops in \textquotedblright
space\textquotedblright\ of their lengths. It is defined as the number of
loops (per unit of volume) with lengths lying between $l$ and $l+dl$. There
are two mechanisms for change of $n(l,t).$ The first one is the mentioned
above deterministic process of evolution of elements of the individual
loops, during which they move undergo the stretching or shrinking. Other
reasons for change of quantity $n(l,t)$ are the random reconnection
processes. We discriminate two types of processes, namely the fusion of two
loops into the larger single loop and the breakdown of a single loop into
two daughter loops (see Fig. 1) \cite{triple}.The kinetics of the vortex
tangle are affected by the intensity of the introduced processes, which is
number of events per unit volume and unit time. The intensity of the first
process is characterized by the coefficient $A(l_{1},l_{2},l)$, which is the
rate of collision of two loops with lengths $l_{1}$and $l_{2}$ and forming
the loop of length $\ l=l_{1}+l_{2}$. The intensity of the second process is
characterized by the coefficient $B(l,l_{1},l_{2})$, which is the rate of
self-intersection and breaking down of a loop with length $l$ into two
daughter loops with lengths $\ l_{1}$ and $l_{2}$. In view of what has been
exposed above we can directly write out the master \textquotedblright
kinetic\textquotedblright\ equation for rate of change of the distribution
function $n(l,t)$
\begin{align}
\frac{\partial n(l,t)}{\partial t}+\frac{\partial n(l,t)}{\partial l}\frac{%
\partial l}{\partial t}& =  \label{kinetic equation} \\
\int \int A(l_{1},l_{2},l)n(l_{1})n(l_{2})\delta
(l-l_{1}-l_{2})dl_{1}dl_{2}\;\;\;\;\;\;\;\;\;\;\;\;l_{1}+l_{2}& \rightarrow l
\notag \\
-\int \int A(l_{1},l,l_{2})\delta
(l_{2}-l_{1}-l)n(l)n(l_{1})dl_{1}dl_{2}\;\;\;\;\;\;\;\;\;\;\;\;\;l_{1}+l&
\rightarrow l_{2}  \notag \\
-\int \int A(l_{2},l,l_{1},)\delta
(l_{1}-l_{2}-l)n(l)n(l_{1})dl_{1}dl_{2}\;\;\;\;\;\;\;\;\;\;\;\;l_{2}+l&
\rightarrow l_{1}  \notag \\
-\int \int B(l,l_{1},l_{2})n(l)\delta
(l-l_{1}-l_{2})dl_{1}dl_{2}\;\;\;\;\;\;\;\;\;\;\;\;\ \;\;\;\;\;\;\;\;\;l&
\rightarrow l_{1}+l_{2}\;\;\;\;\;  \notag \\
+\int \int B(l_{1},l_{2},l)\delta
(l_{1}-l-l_{2})n(l_{1})dl_{1}dl_{2}\;\;\;\;\;\;\;\
\;\;\;\;\;\;\;\;\;\;\;l_{1}& \rightarrow l+l_{2}  \notag \\
+\int \int B(l_{2},l,l_{1})\delta
(l_{2}-l-l_{1})n(l_{2})dl_{1}dl_{2}\;\;\;\;\
\;\;\;\;\;\;\;\;\;\;\;\;\;\;l_{2}& \rightarrow l+l_{1}.  \notag
\end{align}%
All of the processes are depicted at the left of each line. In spite of very
cumbersome form, equation (\ref{kinetic equation}) is quite transparent.
Indeed, let us take for instance the sixth line. It asserts that number of
loops of length $l$ increases whenever a loop with length $l_{1}$ breaks
down into two smaller loops and one of the daughter loops has the length $l$%
. Rate of growth is proportional to number of larger loops $n(l_{1})$ and to
the intensity of breakdown $B(l_{1},l_{2},l\dot{)}$. Then we have to
integrate over all sizes $l_{1}$. Delta function $\delta (l_{1}-l-l_{2})$
just controls conservation of the total length while recombination. We do
not consider here possible small loss of length due to reconnection, this
question in context of our approach had been studied in \cite{Kuzmin}. \
Kinetic\ equation (\ref{kinetic equation}) has a \textquotedblright
book-keeping\textquotedblright\ character, moreover, in this form it is
applicable for other systems e.g., for network of cosmic strings. Physics of
this approach lies in the \textquotedblright correct\textquotedblright\
determinations of coefficient $A(l_{1},l_{2},l)$ and $B(l,l_{1},l_{2})$ of
this equation on the base of some more or less plausible model. In next
subsections we will outline main idea and derive mathematical identities for
the rates coefficient $A(l_{1},l_{2},l)$ and $B(l,l_{1},l_{2})$) in the case
of an arbitrary network of loops. Detailed calculation of these quantities
for vortex loops in the turbulent superfluid helium on the Gaussian model is
performed in Appendix B.

\subsection{Mathematical identities for $A(l_{1},l_{2},l)$ and $%
B(l,l_{1},l_{2}).$}

In this subsection we formulate mathematical definition for quantities $\
A(l_{1},l_{2},l)$ and$\,$\ $B(l,l_{1},l_{2})$. We start with the quantity $%
B(l,l_{1},l_{2}).$ By definition its physical meaning is the frequency of
events when part of line with total length $l$ intersects itself and breaks
down into two daughter loops with lengths $l_{1}$ and $l_{2}$. As it was
already stated we assume that each crossing event leads to the reconnection
of lines. Let us consider function
\begin{equation}
\mathbf{S}_{b}(\xi _{2},\xi _{1},t)=\mathbf{s}(\xi _{2},t)-\mathbf{s}(\xi
_{1},t),  \label{distance_single}
\end{equation}%
which is the vector connecting points $\mathbf{s}(\xi _{2},t_{2})$ and $%
\mathbf{s}(\xi _{1},t_{1})$ (see Fig. 4).
\begin{figure}[tbp]
\includegraphics[width=7cm]{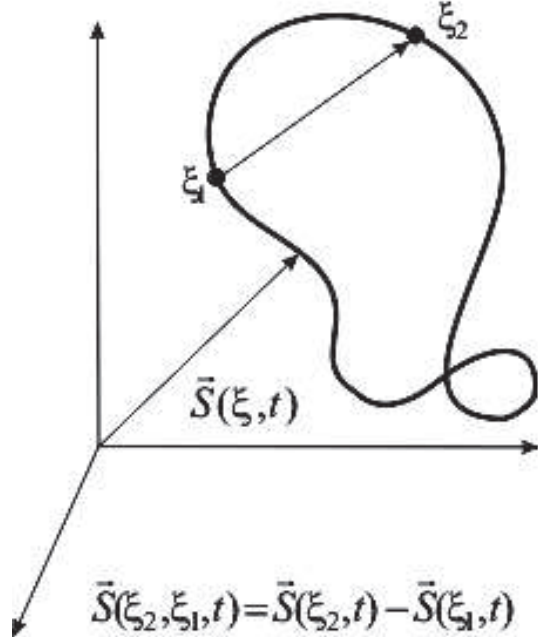}
\caption{Schematic sketch of the self-intersection processes. Elements of
line are described as vectors $\mathbf{s}(\protect\xi )$, where the label
variable $\protect\xi $ is taken here as the arc length. We associate the
moment of intersection with the vanishing of vector $\mathbf{S}(\protect\xi %
_{1},\protect\xi _{1},t)$ connecting points $\mathbf{s}(\protect\xi _{2},t)$
and $\mathbf{s}(\protect\xi _{1},t)$. \ Thus, the rate coefficient \protect%
\smallskip of break-down is equal to number of zeroes of function $\mathbf{S}%
_{b}(\protect\xi _{2},\protect\xi _{1},t)$ in space of its variables $%
\mathbf{\protect\zeta =\{}\protect\xi _{2},\protect\xi _{1},t\}.$}
\end{figure}

Clearly, the condition $\mathbf{S}_{b}(\xi _{2},\xi _{1},t)=0$ implies that
the self-crossing event of parts of the line with label-coordinates $\xi
_{2},\xi _{1}$ occurs at moment of time $t$. \ The quantity $\mathbf{S}%
_{b}(\xi _{2},\xi _{1},t)$ is fluctuating 3-component function of three
arguments $\xi _{2},\xi _{1},t$. We are interested in how often $\mathbf{S}%
_{b}(\xi _{2},\xi _{1},t)$ vanishes in cube of space $\mathbf{\zeta =\{}\xi
_{2},\xi _{1},t\}$. From theory of generalized function it follows that
number of these points (we denote them below as $\mathbf{\zeta }_{a}$) can
be expressed via $\delta $-function of quantity $\mathbf{S}_{b}(\xi _{2},\xi
_{1},t\dot{)}$ with the help of the following formula.
\begin{equation}
\sum_{a}\delta (\mathbf{\zeta }-\mathbf{\zeta }_{a})=\left\vert \frac{%
\partial (X,Y,Z)}{\partial (\xi _{2},\xi _{1},t)}\right\vert _{\mathbf{\zeta
}=\mathbf{\zeta }_{a}}\delta (\mathbf{S}_{b}(\xi _{2},\xi _{1},t)).
\label{s_zeroes_number}
\end{equation}%
Here $X,Y,Z$ are the components of vector $\mathbf{S}_{b}(\xi _{2},\xi
_{1},t)$ . By integration of both parts of (\ref{s_zeroes_number}) over $%
d\xi _{1}d\xi _{2}$ we obtain the full number of intersections (per unit
time). The lengths of pieces of the self-intersecting line are however
arbitrary. The requirement that pieces should have lengths $l_{1}$ and $\
l-l_{1}$ can be satisfied by the introducing additional constraint $\delta
(\xi _{2}-\xi _{1}-l_{1})\,\ $into integrand. In addition we have to do
averaging over all possible fluctuating configurations. Finally the
coefficient $B(l,l_{1},l-l_{1})$ with dimension $\left[ s\right]
=s^{-1}cm^{-1}$ is
\begin{equation}
B(l,l_{1},l-l_{1})=\int \int d\xi _{1}d\xi _{2}\delta (\xi _{2}-\xi
_{1}-l_{1})\,\left\langle \left\vert \frac{\partial (X,Y,Z)}{\partial (\xi
_{2},\xi _{1},t)}\right\vert _{\mathbf{\zeta }=\mathbf{\zeta }_{a}}\delta (%
\mathbf{S}_{b}(\xi _{2},\xi _{1},t))\right\rangle .  \label{B_def}
\end{equation}

To obtain coefficient $A(l_{1},l_{2},l)$ we use the similar procedure. Let
us consider two loops with length $l_{1}$ and $l_{2}$. Our purpose now to
find the rate $A(l_{1},l_{2},l)$ of fusion of these two loops into one loop
of length $l=l_{1}+l_{2}$. Dimension of $A(l_{1},l_{2},l)$ is $%
[A]=cm^{3}s^{-1}$. As previously, we describe vortex filaments by positions
of radius vectors their elements $\mathbf{s}(\xi _{1},t)$ and $\mathbf{s}%
(\xi _{2},t)$. Here we have the two label variables $\xi _{1},\xi _{2}$
belonging to different loops and running in the limits $(0\leq \xi \leq
l_{1})$ and $(0\leq \xi \leq l_{2})$ respectively. One more important
difference with the previous case is that both functions $\mathbf{s}(\xi
_{1},t)$ and $\mathbf{s}(\xi _{2},t)$ should depend on \textquotedblright
initial\textquotedblright\ positions $\mathbf{s}(\xi _{1}=0,t)=\mathbf{R}%
_{1}(t)$ and $\mathbf{s}(\xi _{2}=0,t)=\mathbf{R}_{2}(t)$, chosen arbitrary.
Of course in previous case of the self-intersection of single loop, quantity
$\mathbf{s}(\xi ,t)$ also depended on \textquotedblright
initial\textquotedblright\ positions, but it did not influence the rate of
self-intersection. For case of the fusion this dependance is important,
since very distant loops have the small probability to collide. Let us
introduce the \textquotedblright fusion\textquotedblright\ functions
\begin{equation}
\mathbf{S}_{f}(\xi _{2},\xi _{1},t)=\mathbf{s}(\xi _{2},t)-\mathbf{s}(\xi
_{1},t).  \label{distance_double}
\end{equation}%
Repeating the considerations for case of the single loop we find that the
number of reconnection (per unit of time) of points $\xi _{2},\xi _{1}$
formally coincides with (\ref{s_zeroes_number})
\begin{equation}
\sum \delta (\mathbf{\zeta }-\mathbf{\zeta }_{a})=\left\vert \frac{\partial
(X,Y,Z)}{\partial (\xi _{2},\xi _{1},t)}\right\vert _{\mathbf{\zeta }=%
\mathbf{\zeta }_{a}}\delta (\mathbf{S}_{f}(\xi _{2},\xi _{1},t))
\label{m_zeroes_number}
\end{equation}%
with the difference that $\xi _{2},\xi _{1}$ belong to different curves.
Since intersections of any elements of lines lead to the fusion of loops we
have to integrate (\ref{m_zeroes_number}) over $d\xi _{1}d\xi _{2}$. The
result obtained is valid for chosen pair of loops. To obtain the total
number of events we have to multiply the result obtained by quantity $%
n(l_{1})n(l_{2})d\mathbf{R}_{1}d\mathbf{R}_{2}$, which is the full number of
loops of chosen sizes in the whole volume. Comparing with the master kinetic
equation (\ref{kinetic equation}) we find the final expression for fusion
coefficient $A(l_{1},l_{2},l)$

\begin{equation}
A(l_{1},l_{2},l)=\frac{1}{\mathcal{V}}\int \int d\mathbf{R}_{1}d\mathbf{R}%
_{2}\int \int d\xi _{1}d\xi _{2}\,\left\langle \left\vert \frac{\partial
(X,Y,Z)}{\partial (\xi _{2},\xi _{1},t)}\right\vert _{\mathbf{\zeta }=%
\mathbf{\zeta }_{a}}\delta (\mathbf{S}_{f}(\xi _{2},\xi
_{1},t))\right\rangle ,  \label{A_def}
\end{equation}%
where $\mathcal{V}$ is the total volume of system.

Thus we obtained expressions for the coefficients (\ref{B_def})and (\ref%
{A_def}), which allow to calculate the rates of the fusion and breakdown of
the vortex loops. They are, however, just formal mathematical identities.
Concrete results depend on statistics and dynamics of individual lines.
Therefore to move further we have to ascertain the procedure for averaging.
We will do it with use of the so-called Gaussian model of vortex loops,
which is bases on the presentation them as having a random walk structure.
In order not to overcharge the main text we will expose both ideas of
Gaussian model and detailed evaluation of quantities $B(l_{1},l_{2},l)$ and $%
A(l_{1},l_{2},l)$ in Appendices A and B.

\section{EXACT SOLUTION OF THE "RATE EQUATION"}

\subsection{Zakharov ansatz}

 In this Section we describe one particular but very important
solution to the rate equation (\ref{kinetic equation}). Following \ results
exposed in Appendices we adopt the following expressions for coefficient $A$
and $B$ (see relations (\ref{B_final}),(\ref{A_final}) in Appendix B).
\begin{equation}
A(l_{1},l_{2},l)=b_{m}V_{l}l_{1}l_{2},\;\ \ B(l_{1},l-l_{1},l)=b_{s}\frac{%
V_{l}l}{(\xi _{0}l_{1})^{3/2}}.  \label{A_and_B}
\end{equation}%
Quantity $V_{l}$ $~$ is ($l$-independent) characteristic velocity of
approaching of elements of line, \ $b_{m}$ and $b_{s}$ are numerical
constants. The quantity $\xi _{0}$ associated with the mean radius of the
curvature(see \cite{Nemirovskii_97_1} and Appendix A), and it is a low
cut-off of the whole approach. \

Early the equation similar to (\ref{kinetic equation}) with coefficients $A$
and $B$ (\ref{A_and_B}) had been studied analytically in papers \cite%
{Copeland98}, and numerically in \cite{Steer99}. Thus, in particular in \cite%
{Copeland98} it was demonstrated that (\ref{kinetic equation}) has the
asymptotic solution $n(l)\varpropto e^{-\beta l}l^{-5/2}$, which describes
thermodynamics equilibrium. It had been obtained in supposition of detailed
balance, which implies that each of the line in the collision integral
vanishes. This solution, however, is an approximate solution of the rate
equation (\ref{kinetic equation}) valid only in case of very small daughter
loop.

Here we will search for stationary solution of (\ref{kinetic equation})
neglecting deterministic terms. As it had been discussed in the
Introduction, processes of recombination (fusion and splitting) are the
fastest, so it is quite natural to suppose that the collision term $%
I_{st}\{n\}$ expressed by the lines 2-7 \ in (\ref{kinetic equation}) is the
leading one. We assume that the time independent solution of the equation $%
I_{st}\{n\}=0$ is the basic equation, and nonstationary processes as well as
processes related to the deterministic motion can be accounted in the frame
of the perturbation theory. Therefore, as a first step we neglect other
terms and concentrate on seeking for solution $I_{st}\{n\}=0$.

Coefficients $A(l_{1},l_{2},l)$ and $B(l_{1},l-l_{1},l)$ are the power low
functions, therefore they are scale invariant quantities$.$That implies that
for the scales exceeding $\xi _{0}$ there is no characteristic length in the
statement of problem. It points out that equation $I_{st}\{n\}=0$ should
have the scale invariant, or power-like solution of form $n(l)=C\ast l^{s}$.
To find power-like solution we use the Zakharov ansatz, which \ is the
special treatment of the \textquotedblright collision\textquotedblright\
integral in equation (\ref{kinetic equation}). This trick was elaborated by
Zakharov for the wave turbulence (see e.g., \cite{Zakharov_book}), now we
will show how it works in our case. Let us take for instance the first and
second integrals in the \textquotedblright collision term\textquotedblright\
of (\ref{kinetic equation}). Let us further perform in the second integral
the following change of variables.
\begin{equation}
l=\tilde{l}_{2}\left( \frac{l}{\tilde{l}_{2}}\right) ,\;\ \ \ l_{1}=\tilde{l}%
_{1}\left( \frac{l}{\tilde{l}_{2}}\right) ,\;\;\ \;l_{2}=l\left( \frac{l}{%
\tilde{l}_{2}}\right) .  \label{ansatz}
\end{equation}%
Under this change of variables various factors in the integrand of the
second integral transforms as follows
\begin{equation*}
\delta (l_{2}-l_{1}-l)\rightarrow \left( \frac{l}{\tilde{l}_{2}}\right)
^{-1}\delta (l-\tilde{l}_{1}-\tilde{l}_{2}).
\end{equation*}%
\begin{equation}
n(l)\rightarrow n(\tilde{l}_{2})\left( \frac{l}{\tilde{l}_{2}}\right)
^{s},\;\;\;\;\;n(l_{1})\rightarrow n(\tilde{l}_{1})\left( \frac{l}{\tilde{l}%
_{2}}\right) ^{s},  \label{Zakharov}
\end{equation}%
\begin{equation*}
A(l_{1},l,l_{2})\rightarrow \frac{1}{2}V_{l}\tilde{l}_{1}\tilde{l}_{2}\left(
\frac{l}{\tilde{l}_{2}}\right) ^{2}=A(\tilde{l}_{1},\tilde{l}_{2},l)\left(
\frac{l}{\tilde{l}_{2}}\right) ^{2}.
\end{equation*}%
As result the second integral in the \textquotedblright
collision\textquotedblright\ term takes a form (the additional term $3$ in
the power counting appears from the Jacobian of transformation)
\begin{equation}
\int \int \left( \frac{l}{\tilde{l}_{2}}\right) ^{2+2s-1+3}A(\tilde{l}_{1},%
\tilde{l}_{2},l)n(\tilde{l}_{1})n(\tilde{l}_{2})\delta (l-\tilde{l}_{1}-%
\tilde{l}_{2})d\tilde{l}_{1}d\tilde{l}_{2}\;.\;  \label{ansatz_2}
\end{equation}%
It is easy to see that the transformed second term \ in the
\textquotedblright collision\textquotedblright\ integral in the right hand
side of the master kinetic equation (\ref{kinetic equation}) turns into
first integral with an additional factor $\left( l/\tilde{l}_{2}\right)
^{4+2s}$ in the integrand. Performing the same procedure for all lines we
conclude that the \textquotedblright collision integral\textquotedblright\
of the \textquotedblright rate equation\textquotedblright\ can be written as
\begin{eqnarray}
&&\int \int A(l_{1},l_{2},l)n(l_{1})n(l_{2})\left( 1-\left( \frac{l}{l_{1}}%
\right) ^{4+2s}-\left( \frac{l}{l_{2}}\right) ^{4+2s}\right) \delta
(l-l_{1}-l_{2})dl_{1}dl_{2}\;\;\;\;\;  \label{kin_power} \\
&&-\int \int B(l_{1},l_{2},l)n(l)\left( 1-\left( \frac{l}{l_{1}}\right)
^{s+3/2}-\left( \frac{l}{l_{2}}\right) ^{s+3/2}\right) \delta
(l-l_{1}-l_{2})dl_{1}dl_{2}.\;\;\;\;\;\;\;\;\;\;\;\;\ \;\;  \notag
\end{eqnarray}%
For $s=-5/2$ both expressions $1-\left( \frac{l}{l_{1}}\right)
^{4+2s}-\left( \frac{l}{l_{2}}\right) ^{4+2s}$ and $1-\left( \frac{l}{l_{1}}%
\right) ^{s+3/2}-\left( \frac{l}{l_{2}}\right) ^{s+3/2}$ are equal to $%
(l-l_{1}-l_{2})/l$. Thus the integrands of both integrals in (\ref{kin_power}%
) include expressions of type $\left( x\right) \delta (x)$ and these
integrals vanish. This implies in stationary case and neglecting the
deterministic terms in (\ref{kinetic equation}) the power-like solution $%
n=C\ast l^{-5/2}$ for distribution function $n(l,t)$ of density of loop in
\textquotedblright space\textquotedblright\ of their lengths takes place.%
\newline

\subsection{\protect Flux of length (energy)}

Let us discuss the physical meaning of the solution obtained. First of all
we stress that it is not related to detailed balance i.e. it does not
describe thermal equilibrium, it rather corresponds to the nonequilibrium
state. To clarify the nature of this nonequilibrium state we introduce the
length density $L(t)$ (in space of sizes $l$), which the full length
accumulated in loops of size $l$ (per unit of volume) \cite{length-energy}
\begin{equation}
L(l,t)=n(l,t)l=\frac{\text{the length of all loops with size }l}{\text{unit
of volume*interval of length}}.  \label{dL/dl}
\end{equation}%
The total length (per unit volume), or the vortex line density $\mathcal{L}%
(t)$ is defined as follows:
\begin{equation}
\mathcal{L}(t)=\int L(l,t)dl=\int l\ast n(l,t)dl.  \label{VLD_definition}
\end{equation}%
Quantity $\mathcal{L}(t)$ is obviously conserved during the reconnections
events $\;d\mathcal{L}(t)/dt=0$ (see, however, remark in Section II and
paper \cite{Kuzmin}). Conservation of the vortex line density can be
expressed in the form of continuity equation for the length density $L(l,t)$
\begin{equation}
\frac{\partial L(l,t)}{\partial t}+\frac{\partial P(l)}{\partial l}=0.
\label{continuity equation}
\end{equation}%
This form of equation states that the rate of change of length is associated
with \textquotedblright flux\textquotedblright\ of length in space of sizes
of the loops. Term \textquotedblright flux\textquotedblright\ here means
just the redistribution of length (or energy, see \cite{length-energy})
among the loops of different sizes due to reconnections. Expression for $%
P(l) $ is obtained by multiplying the rate equation \ref{kinetic equation}\
by $l$ and by rewriting the \textquotedblright collision\textquotedblright\
term in the shape of a derivative with respect to $l$. \ The result is
(substitutions $l_{1}/l=x$ and $l_{2}/l=y$ \ have been used below)\newline
\begin{eqnarray}
&&P=\left( \frac{l^{5+2s}}{5+2s}\right) \int \int \frac{1}{2}%
b_{m}V_{l}xyC^{2}x^{s}y^{s}\left( 1-\left( \frac{1}{x}\right) ^{4+2s}-\left(
\frac{1}{y}\right) ^{4+2s}\right) \delta (1-x-y)dxdy  \label{flux_integral}
\\
&&-\left( \frac{l^{s+5/2}}{s+5/2}\right) \int \int \frac{1}{2}b_{s}V_{l}%
\frac{1}{(\xi _{0}x)^{3/2}}C\left( 1-\left( \frac{1}{x}\right)
^{s+3/2}-\left( \frac{1}{y}\right) ^{s+3/2}\right) \delta (1-x-y)dxdy.
\notag
\end{eqnarray}%
Both integrals in relation (\ref{flux_integral}) coincide with integrals in (%
\ref{kin_power}), therefore they vanish for $\ s=-5/2$. However they have
preintegral factors with the denominators, which also vanish for $\ s=-5/2$
and we have the indeterminacy $0/0$ . Calculating numerically integrals in (%
\ref{flux_integral}) as functions of $\ s$\ and taking $s\longrightarrow
-5/2 $ we obtain the final expressions for the \textquotedblright
flux\textquotedblright\ of length in space of sizes of the loops
\begin{equation}
P_{net}=P_{+}-P_{-}=\frac{12.555}{2}C^{2}b_{m}V_{l}-\frac{5.545}{2\xi
_{0}{}^{3/2}}Cb_{s}V_{l}.  \label{Flux}
\end{equation}%
The positive sign of the first term corresponds to the flux of length in the
direction of large scales. This is justified, since the fusion processes
lead to formation of larger and larger loops. The negative sign of the
second term corresponds to the flux of length in the direction of small
scales. This is justified, since the breaking down processes lead to
formation of smaller and smaller loops. Schematically this situation is
depicted in Fig. 5.
\begin{figure}[tbp]
\includegraphics[width=7cm]{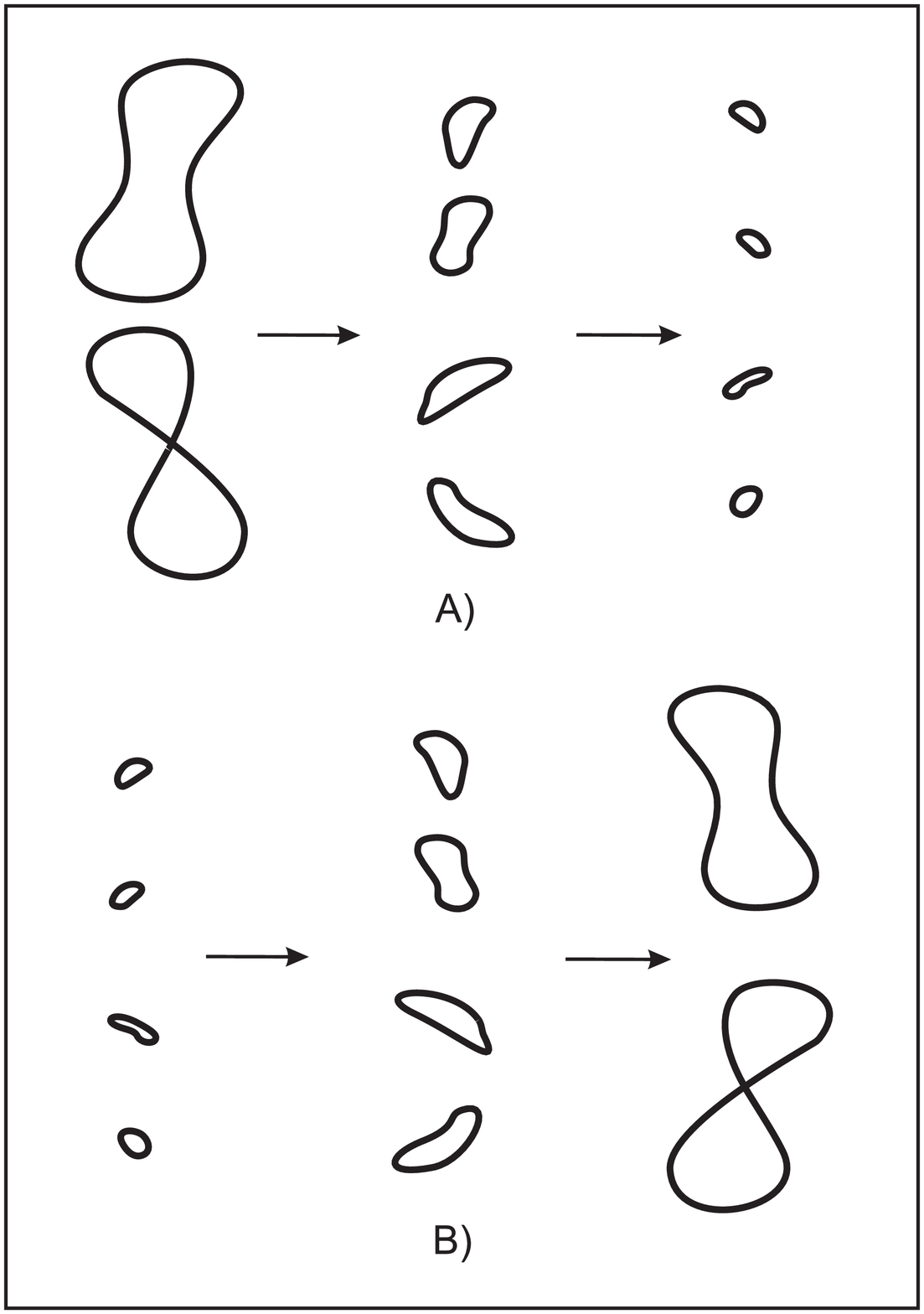}
\caption{Pictures illustrating flux of length (or energy, see \protect\cite%
{length-energy}) in space of the loop sizes. This flux is just
redistribution of total length (energy) among the loops of different sizes
due to recombination process. (a) Negative flux, or direct cascade appears
due to consequent break-down resulting in formation of smaller and smaller
loops. (b). Positive flux, or inverse cascade describes consequent fusion of
loops leading to formation of larger and larger loops.}
\end{figure}
We will also use terms "direct cascade" describing cascade like breakdown of
loops and "inverse cascade" responsible for formation of larger and larger
loops. As it had been discussed in Introduction the direct cascade of
consequent breaking down of vortex loops was predicted by Feynman in his
pioneering paper \cite{Feynman}. The Feynman's idea was confirmed in various
numerical calculations, where the procedure of artificial elimination of
small loops had been used \cite{Schwarz88}-\cite{Tsubota00}. Our analytical
calculations confirmed the splendid Feynman's conjecture and give an exact
evaluation for the cascade-like flux (the second term in the right hand side
of relation (\ref{Flux})). In addition we obtained result about inverse
cascade responsible for formation of larger and larger loops, so the
direction of the net flux is not clear. In more details this fact will be
discussed in the next Section.

Thus we have found stationary power-solution to the rate equation (\ref%
{kinetic equation}) neglecting deterministic terms. As \ it was already
mentioned this solution connected with recombination of loops describes the
fastest processes \ and can be considered as a first iteration for the whole
problem stated by master "rate equation" (\ref{kinetic equation}). The
approach elaborated above allows to draw several conclusions concerning both
the structure and dynamics of the real vortex tangle in the turbulent He II.
It will be done in the following Section.

\section{ON THEORY OF SUPERFLUID TURBULENCE.}

\subsection{\protect Properties of the vortex tangle with no normal
component (zero temperature case).}

In this subsection we discuss some properties of the vortex tangle resulting
from the solution of equation $I_{st}\{n\}=0$ obtained and analyzed in the
previous Section. As mentioned this solution is a stationary solution of the
master kinetic equation (\ref{kinetic equation}) neglecting deterministic
terms. The latter implies that the interaction with normal component is
omitted hence the title of this subsection.

To use formulas derived in previous section we have to specify quantity $%
V_{l}$, which enters into the rates coefficients $A$ and $B$ of both the \
merging and breaking down processes. Basing on the results obtained in
Appendix B we estimate the velocity factor $V_{l}$ to be of the order of $%
\sqrt{2}\kappa /\xi _{0}$ \ ($\kappa $ is the quantum of circulation). Thus
the only parameters of the whole theory (at zero temperature) are the
quantum of circulation $\kappa $\ and the mean radius of curvature $\xi _{0}$%
.

\subsubsection{Vortex Line Density and the mean curvature.{\protect\large \ }}

 Because of a huge amount of reconnections each of the terms in the
right hand side of relation for the net flux (\ref{Flux}) are large. The net
flux $P_{net}$, which is the difference between positive $P_{+}$ and
negative $P_{-}$ constituents is much smaller. Neglecting $P_{net}$ and
equating $P_{+}$ and $P_{-}$ we are in position to ascertain constant $C$
\begin{equation}
C=\frac{5.455}{12.555}\frac{b_{s}}{b_{m}}\frac{1}{\xi _{0}^{3/2}}=C_{VLD}%
\frac{1}{\xi _{0}^{3/2}}.  \label{C_constant}
\end{equation}

 New numerical parameter $C_{VLD}\approx $ $1.\,\allowbreak
810\,4\times 10^{-2}$. Thus the power-like solution $n(l)$ of the master
"rate equation" (\ref{kinetic equation}) is
\begin{equation}
n(l)=\frac{C_{VLD}}{\xi _{0}^{3/2}}l^{-5/2}.  \label{n_final}
\end{equation}%
The total length \ $\mathcal{L}$ per unit of volume is evaluated as follows
(we recall that quantity $\xi _{0}$ serves as the low cut-off ):
\begin{equation}
\mathcal{L=}\int_{\xi _{0}}^{\infty }l\ast n(l)dl=\frac{2C_{VLD}}{\xi
_{0}^{2}}.  \label{VLD_vs_curvature}
\end{equation}

 Result (\ref{VLD_vs_curvature}) is remarkable. It asserts that
interline space $\delta =\mathcal{L}^{-1/2}$ is of the order of the mean
radius of curvature $\xi _{0}$, namely
\begin{equation}
\xi _{0}=\sqrt{2C_{VLD}}\mathcal{L}^{-1/2}  \label{curvature_vs_interline}
\end{equation}

\begin{figure}[tbp]
\includegraphics[width=7cm]{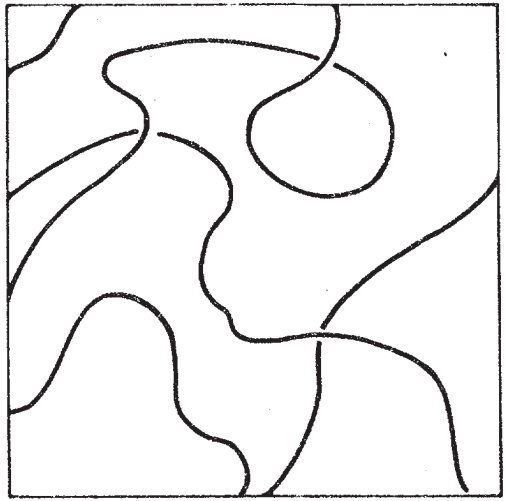}
\caption{Likely configuration of the vortex tangle when mean radius of
curvature is of order of interline space.}
\end{figure}
This idea had been launched by Schwarz \cite{Schwarz88}, it is illustrated
in Fig. 6. The nature of this phenomenon was not clear. We proved that this
relation appears due to kinetics of colliding vortex loops. In more
realistic situation of nonzero temperature connection between interline
space the mean radius of curvature $\xi _{0}$ had been obtain numerically by
Schwarz. \cite{Schwarz88}
\begin{equation}
\xi _{0}=\frac{1}{c_{2}(T)\sqrt{2}}\mathcal{L}^{-1/2}.
\label{Schwarz_number}
\end{equation}%
The temperature dependent parameter $c_{2}(T)$ is one of the structure
constant of the voretx tangle introduced by Schwarz. It is responsible for
kinking of the vortex filaments. Function $c_{2}(T)$ decreases when the
temperature grows, this imply that the vortex tangle becomes more kinked at
low temperatures. This fact had been reported in numerical works (see e.g.,
\cite{Schwarz88},\cite{Tsubota00}). Comparing (\ref{Schwarz_number}) \ and (%
\ref{curvature_vs_interline}) we conclude that parameter $C_{VLD}$, obtained
in our approach is the zero-temperature limit of quantay $1/4c_{2}^{2}(T).$%
For the minimal temperature $T=1.07\ K$ covered in the Schwarz's simulations
the latter quantity is approximately $2.\,\allowbreak 04\times 10^{-2}$ .
Note that this value is very close to $C_{VLD}\approx $ $1.\,\allowbreak
810\,4\times 10^{-2}$, obtained for the zero temperature case.

\subsubsection{\protect \qquad Net flux and vortex line density}

 In previous subsection we had ascertained the constant $C$ in
relation $n(l)=C\ast l^{s}$ (see (\ref{n_final})) and found the connection
between the vortex line density $\mathcal{L}$ with mean curvature $\xi _{0}$
(see (\ref{VLD_vs_curvature})). This enables us to express the net flux $%
P_{net}$ (\ref{Flux}) via quantity $\mathcal{L}$. Substituting (\ref%
{VLD_vs_curvature}) into relation for the net flux (\ref{Flux}) we arrive at
conclusion that both the positive constituent $P_{+}$ and the negative one $%
P_{-}$ are proportional to the squared vortex line density $\mathcal{L}^{2}.$
In unsteady case at finite temperature quantities $P_{+}$ and $P_{-}$ do not
compensate each other, so the net flux $P_{net}$ does not vanish and it is
also proportional to the squared vortex line density $\mathcal{L}^{2}.$ That
means that the rate of decay of quantity $\mathcal{L}(t)$ due to fluxes
carrying away the length from the system can be written as
\begin{equation}
\;\;\;\frac{d\mathcal{L}(t)}{dt}\propto -\mathcal{L}^{2}\;.\;  \label{VE}
\end{equation}%
Relation (\ref{VE}) is the particular case of the so-called Vinen equation
discussed in detail in the next Subsection. It is remarkable that relation (%
\ref{VE}) appears due to the reconnection processes. \ The own dynamics of
filament specific for various systems is absorbed by the \ quantity $\xi
_{0},$ which has dropped out of the Vinen equation at all. Thus the (\ref{VE}%
) has the universal character and can be applied for other systems. It
reflects growth $\delta =\mathcal{L}^{-1/2}\propto \sqrt{t}$ of the
interdefects space, which is general behavior for nonconserved order
parameter (see \cite{Rivers}).The result obtained requires one comment. We
used stationary solution of the rate equation to describe unsteady
situation. This can be justified only when change of $\mathcal{L}(t)$ is
slow and structure of loops (namely quantity $\xi _{0}$) has a time to
adjust its equilibrium value expressed by (\ref{VLD_vs_curvature}). This was
confirmed in numerical simulations in \cite{S_Rozen}. Resuming result of
this Subsubsection we would like to stress that our calculations confirmed
the Feynman's conjecture on the formation of cascade-like breakdown of
vortex loops, leading to decay of the vortex tangle.

\subsubsection{Full rate of reconnections.}

The full rate of reconnection $\dot{N}_{rec}$ can be evaluated directly from
collision term in the master \textquotedblright rate
equation\textquotedblright\ (\ref{kinetic equation}). Indeed, this term
describes change of $n(l)$ due to reconnection events. It takes into account
signs of events, depending on whether the loop of size $l$ appears or dies
in result of reconnection. Therefore, if we take all terms in collision
integral with the plus sign and use for estimation our solution for $n(l)$
we obtain the total number of reconnections. The according calculations lead
to this result
\begin{equation*}
\dot{N}_{rec}=\frac{1}{3}\frac{\kappa (b_{s}C_{VLD}+b_{m}^{2}C_{VLD})}{\xi
_{0}^{5}}=C_{rec}\kappa \mathcal{L}^{5/2},
\end{equation*}%
where $C_{rec}$ one more constant of the order $0.1-0.5.$ This result agrees
with the recent numerical investigation\cite{Barenghi2004}.\newline

\subsection{Vinen equation}

The aim of this subsection is to study one of the key questions of the
theory of superfluid turbulence, namely the evolution of the vortex line
density defined in relation (\ref{VLD_definition}). Unlike the previous
subsection we do not omit deterministic terms in (\ref{kinetic equation}),
which implies that the interaction with the normal component is taken into
consideration. Let us multiply the kinetic equation (\ref{kinetic equation})
by$\ l$ and integrate over all sizes.
\begin{equation}
\frac{d\mathcal{L}(t)}{dt}=\int \frac{\partial n(l,t)}{\partial t}ldl=-\int
\frac{\partial n(l,t)}{\partial l}\frac{\partial l}{\partial t}%
ldl-\left\vert P_{net}\right\vert .\;\;  \label{VE_0}
\end{equation}%
The first term in the right hand side of (\ref{VE_0}) describes a change of
vortex line density $\mathcal{L}(t)$ due to the deterministic motion, in
fact due to the mutual friction. \ Quantity $P_{net}$ is the net
\textquotedblright flux\textquotedblright\ of the length (or energy, see
subsection II. B) in $l-$space. We use the absolute value of $\ \left\vert
P_{net}\right\vert $ because the net flux \ $P_{net}$ always carries away
the vortex line density $\mathcal{L}$ \ from the system, and different signs
refer to direction of the cascade.

Let us treat the deterministic term in equation (\ref{VE_0}). We calculate
the rate of a change in the length of each loop on the base of the motion
equation of \ the line in the so-called local approximation (see e.g., \cite%
{Schwarz88}). In this approach the velocity of the line element $\mathbf{v}%
_{l}(\xi )$ is
\begin{equation}
\mathbf{v}_{l}=\beta \mathbf{s}^{\prime }\times \mathbf{s}^{\prime \prime
}+\alpha \mathbf{s}^{\prime }\times (\mathbf{V}_{ns}-\beta \mathbf{s}%
^{\prime }\times \mathbf{s}^{\prime \prime })+\alpha ^{\prime }\mathbf{s}%
^{\prime }\times \mathbf{s}^{\prime }\times (\mathbf{V}_{ns}-\beta \mathbf{s}%
^{\prime }\times \mathbf{s}^{\prime \prime }).  \label{line_velocity}
\end{equation}%
Here $\mathbf{s}^{\prime }$ and $\mathbf{s}^{\prime \prime }$ are the first
and second derivatives from position of line $\mathbf{s}(\xi )$ with respect
to label variable $\xi $, which coincides here with the arc length.
Quantaties $\alpha $ and $\alpha ^{\prime }$ are the temperature dependent
friction coefficients. To calculate $\partial l/\partial t,$ we use the
relation for the rate of a change in the length $\partial \delta l/\partial
t $ \ for some arbitrary element with length $\delta l$. Assuming for a
while that the label variable $\xi $ is not exactly the arc length, we have $%
\delta l=\left\vert \mathbf{s}^{\prime }\right\vert $ $\delta \xi .~\ $Then\
the \ following chain of relations takes place%
\begin{equation}
\frac{\partial \delta l}{\partial t}=\frac{\partial \left\vert \mathbf{s}%
^{\prime }\right\vert \delta \xi }{\partial t}=\frac{\left\vert \mathbf{s}%
^{\prime }\right\vert }{\left\vert \mathbf{s}^{\prime }\right\vert }\frac{%
\partial \left\vert \mathbf{s}^{\prime }\right\vert \delta \xi }{\partial t}=%
\frac{\mathbf{s}^{\prime }}{\left\vert \mathbf{s}^{\prime }\right\vert }%
\frac{\partial \mathbf{s}^{\prime }\delta \xi }{\partial t}=\mathbf{s}%
^{\prime }\mathbf{v}_{l}^{\prime }\delta \xi .  \label{ddl/dt}
\end{equation}%
On the last stage we return to condition $\left\vert \mathbf{s}^{\prime
}\right\vert =1.$ Differentiating (\ref{line_velocity}) and multiplying by $%
\mathbf{s}^{\prime }$ we have after little algebra
\begin{equation}
\frac{\partial \delta l}{\partial t}=(\alpha (\mathbf{s}^{\prime }\times
\mathbf{s}^{\prime \prime })\mathbf{V}_{ns}-\alpha \beta (\mathbf{s}^{\prime
}\times \mathbf{s}^{\prime \prime })^{2})\delta \xi .  \label{growth_dl}
\end{equation}%
Terms with $\alpha ^{\prime }$ vanish due to symmetry. The next step is to
average expression (\ref{ddl/dt}) over all possible configurations of the
vortex loops. We do it with use of the Gaussian model of the vortex tangle%
\cite{anisotropy}. In accordance with this model
\begin{equation}
\langle \mathbf{s}^{\prime }\times \mathbf{s}^{\prime \prime }\rangle \;=%
\frac{I_{l}}{\sqrt{2}c_{2}\xi _{0}}\frac{\mathbf{V}_{ns}}{\left\vert \mathbf{%
V}_{ns}\right\vert },\ \ \ \ \langle (\mathbf{s}^{\prime }\times \mathbf{s}%
^{\prime \prime })^{2}\rangle \;=\langle (\mathbf{s}^{\prime \prime
})^{2}\rangle =\frac{1}{2\xi _{0}^{2}}.\ \ \ \ \   \label{s's''}
\end{equation}%
Quantity $I_{l}$ is another (together with $c_{2}$) structure constant
introduced by Schwarz \ \cite{Schwarz88}). Substituting (\ref{s's''}) into
(averaged equation (\ref{ddl/dt})) and then into (\ref{VE_0}) and
integrating by part we get the contribution into $d\mathcal{L}(t)/dt\ $from
the deterministic term
\begin{equation}
(\alpha \frac{I_{l}\left\vert \mathbf{V}_{ns}\right\vert }{\sqrt{2}c_{2}\xi
_{0}}-\alpha \beta \frac{1}{2\xi _{0}^{2}})\int \frac{\partial n(l,t)}{%
\partial l}l^{2}dl=-\alpha \frac{2I_{l}\left\vert \mathbf{V}_{ns}\right\vert
}{\sqrt{2}c_{2}\xi _{0}}\mathcal{L+}\frac{\alpha \beta }{\xi _{0}^{2}}%
\mathcal{L}.  \label{VE_det}
\end{equation}

Now we have to treat the "flux" term in equation (\ref{VE_0}). We consider
consequently the collision and reconnection events\ to put the system into
the equilibrium (with respect to solution (\ref{n_final})) state much faster
than the slow deterministic processes. This implies that the parameters $\xi
_{0}$, $c_{2}$ and $I_{l}$ have a time to adjust to their equilibrium
values. This assumption is widely adopted and it was confirmed in numerical
simulations\cite{S_Rozen}. By use of expression for the net flux(\ref{Flux}%
), ridding of the constant $C$ with the help of the normalization condition $%
\mathcal{L}(t)=\int n(l)ldl$ and using definition of the Schwarz number $%
c_{2}(T)$ (see Subsussection IV.A.1, relations (\ref{VLD_vs_curvature}),(\ref%
{Schwarz_number})) we rewrite the expression for flux (\ref{Flux}) in form $%
P_{net}=C_{F}\kappa \mathcal{L}^{2},$ where the temperature constant $C_{F}$
is
\begin{equation}
C_{F}\approx (2.22b_{m}-3.92\,6c_{2}^{2}\allowbreak b_{s}).
\label{Feynman constant}
\end{equation}

    We named this constant in honor of Feynman who was the first person
to discuss evolution of vortex line density due to the reconnection
processes. We would like to recall that Feynman supposed the decay of a
vortex tangle due to the cascade-like breakdown of vortex loops with further
disappearance of them on very small scales. The approach elaborated here
quantitatively confirmed Feynman's splendid conjecture. Relation (\ref%
{Feynman constant}) shows, however that there is also possible the inverse
cascade, which corresponds to the cascade-like fusion of \ vortex loops.
Unfortunately our approach has too approximate character to do any strong
quantitative conclusion. It is clear, however, that for low temperatures,
where the vortex tangle is more kinked, correspondingly $c_{2}$ is large,
the quantity $C_{F}$ \ is negative. This corresponds to the direct cascade
in region of very small loops. On the contrary for \ high temperature lines
are smoother, $c_{2}$ is small, and $C_{F}$ \ is positive, which implies
that there is inverse cascade with formation of large loops. If we for
instance adopt values for $b_{m}$ and $b_{s}$ and use for $c_{2}^{2}$ values
offered by Schwarz (see \cite{Schwarz88}), then we get $C_{F}\ \approx
-0.252 $ for the temperature $1.07$ K and $\ C_{F}\approx 0.4$ K for the
temperature $2.01$ K.

Collecting contribution into $d\mathcal{L}(t)/dt$ from both the
deterministic and collision processes (\ref{VE_0}), (\ref{VE_det}) and (\ref%
{Feynman constant}), and taking into account that $P_{net}=\left\vert
C_{F}\right\vert \kappa \allowbreak \mathcal{L}^{2}$ we finally have%
\begin{equation}
\frac{d\mathcal{L}(t)}{dt}=\frac{5}{2}\alpha I_{l}\left\vert
V_{ns}\right\vert \mathcal{L}^{\frac{3}{2}}-\frac{5}{2}\alpha \beta c_{2}^{2}%
\mathcal{L}^{2}-\left\vert C_{F}\right\vert \kappa \allowbreak \mathcal{L}%
^{2}.  \label{Vinen}
\end{equation}%
Thus, starting with kinetics of a network of vortex loops, we get the famous
Vinen equation\cite{Vinen}.

Let us discuss the meaning of various terms entering this equation. The
first, generating term in the right hand side of the Vinen equation
describes the grows of the vortex tangle due to the mutual friction. The
second term is also connected to the mutual friction, however this term is
responsible for a decrease of the vortex line density. This point of view
coincides with ideas by Schwarz \cite{Schwarz88} who obtained the
deterministic contribution into $d\mathcal{L}(t)/dt$ using a bit different
approach. The third term in the right hand side of (\ref{Vinen}) is related
to the random collisions of vortex loops. It describes a decrease of the
vortex line density due to the flux of length carrying away the length from
the system. Depending on an interplay between coefficients $b_{m}$ and $%
b_{s} $ and the Schwarz parameters $c_{2}(T)$ \ the flux can be either
positive or negative. We stress again that independently on the sign of the
net flux, this third term should result in a decrease of the vortex line
density. \ The negative flux appears when the break down of loops prevails
and the cascade-like process of generation of smaller and smaller loops
forms. There exists a number of mechanisms of disappearance of the vortex
energy on very small scales. It can be e.g., the acoustic radiation, collapse
of lines, Kelvin waves etc. These dissipative mechanisms balance the grow of
the line length due to the mutual friction. As a result, fully developed
turbulence with the flux of energy in direction of small scales is formed,
what implies highly chaotic picture of the vortex tangle (see Fig. 7).
\begin{figure}[tbp]
\includegraphics[width=7cm]{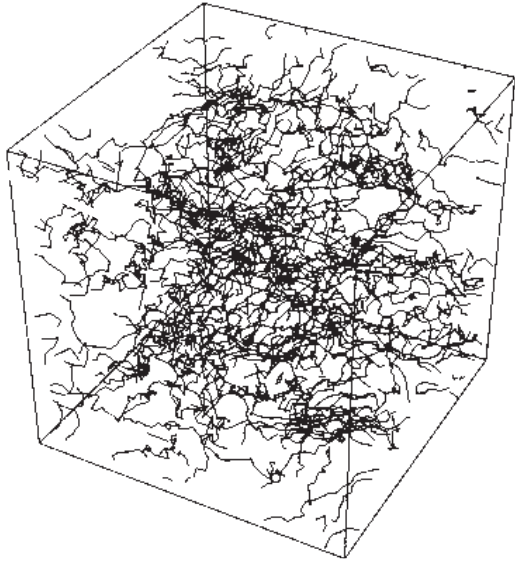}
\caption{The low temperature numerical simulation of the vortex tangle%
\protect\cite{Tsubota00}. The ensemble of vortex loops develops into a
highly chaotic regime. }
\end{figure}

The case with inverse is less clear. The inverse cascade implies the
cascade-like process of generation of larger and larger loops. Unlike the
previous case of the direct cascade, there is no an apparent mechanism for
disappearance of very large loops. The probable scenario is that the parts
of large loops are pinned to the walls. Finally, a state with few lines
stretching from wall to wall with poor dynamics and rare events is realized,
this is a degenerated state of the vortex tangle (See Fig. 8).
\begin{figure}[tbp]
\includegraphics[width=7cm]{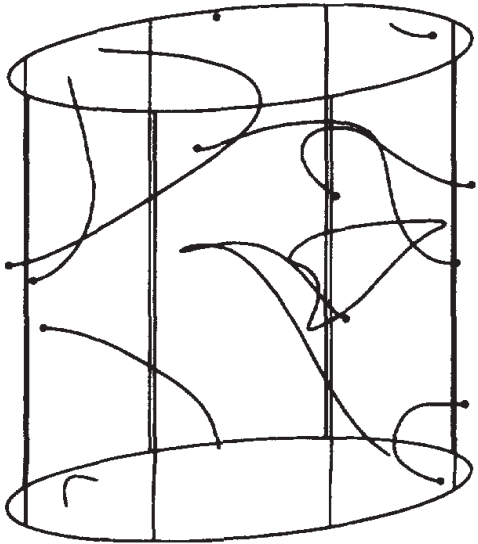}
\caption{The high temperature numerical simulation of the vortex tangle%
\protect\cite{Aarts}. It is seen that vortex tangle is generated into the
state with very few lines.}
\end{figure}
Some of numerical investigators\cite{Schwarz88}, \cite{Aarts}, report on
this situation. This observation can be an alternative explanation for a
phenomenon discovered in Helsinki group\cite{Finne2003}, who observed
transition to superfluid turbulence governed by the temperature.

\section{SUMMARY AND CONCLUSION}

The evolution of a network of vortex loops in He II, which merge and break
down due to reconnections has been considered. It was discussed that because
of very frequent reconnections of the vortex loops these processes of
recombination is the leading mechanism in the whole dynamics of the vortex
tangle. To develop the quantitative description we take the vortex loops to
have a Brownian structure with the only degree \ of freedom, which is length
$l$ of the loop. We perform investigation on the basis of the Boltzmann type
\textquotedblright kinetic equation\textquotedblright\ for distribution
function $n(l)$ of number of loops with lengths $l.$ By the use of special
substitution of variables in the collision integral (Zakharov ansatz) we had
found the power-like stationary solution to this equation. This is a
non-equilibrium solution characterized by two mutual fluxes of length
(energy) in the space of loop sizes. The negative flux (direct cascade)
corresponds to the cascade-like breaking down the vortex loops with
consequent dissipation of energy on a very small scale. This situation fully
coincides with the scenario of superfluid turbulence proposed by Feynman\cite%
{Feynman}. The positive flux (inverse cascade) corresponds to the
cascade-like formation of larger and larger loops. Analyzing this
solution we drew several results on the structure and dynamics of the vortex
tangle in the superfluid turbulent helium. In particular, we obtained that
the mean radius of the curvature is of the order of the interline space. We
also evaluated the full rate of reconnection events. Assuming, further, that
processes of random colliding are the fastest we studied evolution of \ the
vortex line density $\mathcal{L}(t)$ in a presence of mutual friction (for
finite temperatures). This evolution was shown to obey the famous Vinen
equation. In conclusion we discuss the properties of the Vinen equation from
the point of view of the developed approach. Thus, depending on the
temperature (and independently on velocity) vortices either develop into a
highly chaotic picture (turbulence), or degenerate into few smooth lines.

\begin{center}
\textbf{ACKNOWLEDGMENTS}
\end{center}

This work was partially supported by grants N 05-08-01375 and 07-02-01124
from the RFBR and grant of President Federation on the state support of
leading scientific schools RF NSH-6749.2006.8. \ I am grateful to
participants of the workshop \textquotedblright Superfluidity under
Rotation\textquotedblright\ (Jerusalem, 2007) for useful discussion of the
results exposed above.

\renewcommand{\theequation}{A.\arabic{equation}}
\setcounter{equation}{0} \appendix

\section*{APPENDIX A: GAUSSIAN MODEL}

To evaluate quantities $B(l_{1},l_{2},l)$ and $A(l_{1},l_{2},l)$ written in
form (\ref{B_def}) and (\ref{A_def}) one needs to know statistics of
individual loops. In general, this statistics should be extracted from an
investigation of the full dynamical problem. The according statement of such
problem includes equation of the motion (Biot-Savart law for quantum
vortices) and dissipative effects (interaction with normal component). This
problem is very involved, and at this stage we choose another way, namely,
we use the Gaussian model of the vortex tangle elaborated by author \cite%
{Nemirovskii_97_1}. Gaussian model uses the supposition that vortex loops
have a random walk (or Brownian) structure. Main mathematical tool
to describe the random walk structure is the Wiener distribution (see e.g.,
\cite{Kleinert},\cite{Wiegel73}). The pure Wiener distribution has some
deficiencies to describe real vortex filament. The most apparent one is that
Wiener distribution does not have finite average $\left\langle \mathbf{s}%
^{\prime }(\xi ,t)\mathbf{s}^{\prime }(\xi ,t)\right\rangle $, which is a
squared tangent vector. Moreover it does not have the squared second
derivative $\left\langle \mathbf{s}^{\prime \prime }(\xi ,t)\mathbf{s}%
^{\prime \prime }(\xi ,t)\right\rangle $, which is a squared curvature
vector.$.$ In classical form it also does not describe possible anisotropy
and polarization of the loops. To overcome these difficulties the so-called
generalized Wiener distribution had been offered~in paper \cite%
{Nemirovskii_97_1} . The generalized Wiener distribution allows to take into
account the possible anisotropy and finite curvature. Namely, the
probability $\mathcal{P}(\left\{ \mathbf{s}(\xi ,t)\right\} )$\ to find some
particular configuration $\left\{ \mathbf{s}(\xi ,t)\right\} $ is expressed
by the probability distribution functional (see for details the paper by
author \cite{Nemirovskii_97_1})
\begin{equation}
\mathcal{P}(\left\{ \mathbf{s}(\xi ,t)\right\} )=\mathcal{N}\exp \left(
-\int\limits_{0}^{l}\int\limits_{0}^{l}\mathbf{s}^{\prime \alpha }(\xi
_{1},t)\Lambda ^{\alpha \beta }(\xi _{1}-\xi _{2})\mathbf{s}^{\prime \beta
}(\xi _{2},t)d\xi _{1}d\xi _{2}\right) .  \label{Gauss_model}
\end{equation}%
Here $\mathcal{N}$ is normalizing factor, $l$ is the length of curve.
Parameters of this generalized Wiener distribution (elements of matrix $%
\Lambda ^{\alpha \beta }(\xi _{1}-\xi _{2})$) were taken so that some
quantities (e.g., mean curvature, coefficients of anisotropy, etc.)
evaluated on the basis of (\ref{Gauss_model}) give the values known from
both experimental studies and numerical simulations. Typical form of
function $\Lambda _{\alpha \beta }(\xi ,\xi ^{\prime })$ is a smoothed $%
\delta $ function of a Mexican hat shape\ with the width equal $\xi _{0}.$
According to this model the \textquotedblright average\textquotedblright\
vortex loop has a typical structure shown in Fig. 9.
\begin{figure}[tbp]
\includegraphics[width=7cm]{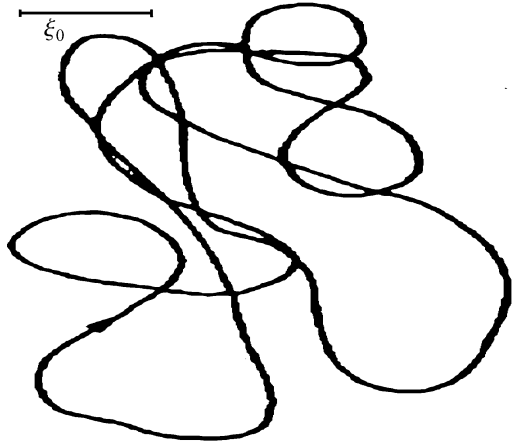}
\caption{Snapshot of the \textquotedblright average\textquotedblright\
vortex loop obtained from analysis of the statistical properties. Close $%
(\Delta \protect\xi \ll \protect\xi _{0})$ parts of the line are separated
in $3D$ space by distance $\Delta \protect\xi $. The distant parts $(\protect%
\xi _{0}\ll \Delta \protect\xi )$ are separated in $3D$ space by the
distance $\protect\sqrt{\protect\xi _{0}\Delta \protect\xi }$, i.e. the
vortex loop has the typical random walk structure. The scale $\protect\xi %
_{0}$ is depicted here in the left upper corner.}
\end{figure}
The average loop can be imagined as consisting of many arches with the mean
radius of curvature equal $\xi _{0}$ randomly (but smoothly) connected to
each other. The close parts of the loop separated (along line) by distance $%
\xi _{2}-\xi _{1}$ smaller then the mean radius of curvature $\xi _{0}$ are
strongly correlated, $\left\langle \mathbf{s}^{\prime }(\xi _{1},t)\mathbf{s}%
^{\prime }(\xi _{2},t)\right\rangle \rightarrow 1,$ ($\mathbf{s}^{\prime }$
is the tangent vector) and line is smooth. \ Remote parts of the line $\ \
(\xi _{2}-\xi _{1}\gg \xi _{0})$ are not correlated at all, $\left\langle
\mathbf{s}^{\prime }(\xi _{1},t)\mathbf{s}^{\prime }(\xi
_{2},t)\right\rangle \rightarrow 0$. Thus for large separations the vortex
loop has a typical "random walk" structure. This \textquotedblright
semifractal\textquotedblright\ behavior satisfies to the generalized Wiener
distribution.

    Being the Gaussian, the Wiener distribution allows to calculate
readily any average functional $A(\left\{ \mathbf{s}(\xi ,t)\right\} )$
depending on configuration $\left\{ \mathbf{s}(\xi ,t)\right\} .$It can be
done evaluating the following path integral%
\begin{equation*}
\left\langle \mathcal{A}(\left\{ \mathbf{s}(\xi ,t)\right\} )\right\rangle {%
\ =}\int DsA(\left\{ \mathbf{s}(\xi ,t)\right\} )\mathcal{P}(\left\{ \mathbf{%
s}(\xi ,t)\right\} ).
\end{equation*}

In practice it is more convenient to deal with the characteristic functional
$W(\{\mathbf{P}(\xi ,t)\})$ defined as
\begin{equation}
W(\{\mathbf{P}(\xi ,t)\})\;=\left\langle \exp \left( i\int\limits_{0}^{l}%
\mathbf{P}(\xi ,t)\mathbf{s}^{\prime }(\xi ,t)d\xi \ \right) \right\rangle
\;.  \label{CF_def}
\end{equation}%
The characteristic functional enables us calculate any averages depending on
vortex lines configuration $\left\{ \mathbf{s}(\xi ,t)\right\} $ by a simple
functional differentiation. For instance, the average tangent vector $%
\left\langle \mathbf{s}_{\alpha }^{\prime }(\xi _{1})\right\rangle $, or the
correlation function between orientation of the different elements of the
vortex filaments $\left\langle \mathbf{s}_{\alpha }^{\prime }(\xi _{1})%
\mathbf{s}_{\beta }^{\prime }(\xi _{2})\right\rangle $ are readily expressed
via the characteristic functional accordingly to the following rules:

\begin{equation}
\left\langle \mathbf{s}_{\alpha }^{\prime }(\xi _{1})\right\rangle =\;\left.
\frac{\delta W}{i\delta \mathbf{P}^{\alpha }(\xi _{1})}\right\vert _{\;%
\mathbf{P}\;=\;0},\;\;\;\;\;\left\langle \mathbf{s}_{\alpha }^{\prime }(\xi
_{1})\mathbf{s}_{\beta }^{\prime }(\xi _{2})\right\rangle =\;\left. \frac{%
\delta ^{2}W}{i\delta \mathbf{P}^{\alpha }(\xi _{1})\;\;i\delta \mathbf{P}%
^{\beta }(\xi _{2})}\right\vert _{\mathbf{P}\;=\;0}  \label{r'(W)}
\end{equation}%
Calculation of the characteristic functional $W(\{\mathbf{P}(\xi ,t)\})$ (%
\ref{CF_def}) on the base of the probability functional \ (\ref{Gauss_model}%
) is reduced to the functional integration, which, in turn, reduces to the
\textquotedblright full square procedure\textquotedblright . \ The result is

\begin{equation}
W(\{\mathbf{P}(\xi,t)\})\;=\exp \left\{
-\int\limits_{0}^{l}\int\limits_{0}^{l}\mathbf{P}^{\alpha }(\xi
_{1})N^{\alpha \beta }(\xi _{1}-\xi _{2})\mathbf{P}^{\beta }(\xi _{2})d\xi
_{1}d\xi _{2}\right\}.  \label{CF_N}
\end{equation}

At this stage, to avoid lengthy calculations we simplify the model expressed
by the probability distribution functional (\ref{Gauss_model}), namely we
omit both the anisotropy and \ polarization. We also disregard the closure
condition of lines, which is not significant for the rate coefficients. In
this case the matrix $N^{\alpha \beta }(\xi _{1}-\xi _{2})$ used in (\cite%
{Nemirovskii_97_1}) is proportional to unit matrix and has to be can be
taken as
\begin{equation}
N^{\alpha \beta }(\xi _{1}-\xi _{2})=\frac{\delta _{\alpha \beta }}{6}\exp %
\left[ -\frac{(\xi _{1}-\xi _{2})^{2}}{4\xi _{0}^{2}}\right] .
\label{N_isotrop}
\end{equation}

    For small separation $\ (\xi _{2}-\xi _{1}\ll \xi _{0})$ the sum $%
\sum\nolimits_{\alpha }N^{\alpha \alpha }(\xi _{1}-\xi _{2})\rightarrow 1/2$%
, that guarantees that $\left\langle \mathbf{s}^{\prime }(\xi ,t)\mathbf{s}%
^{\prime }(\xi ,t)\right\rangle =1,$ as it should be for smooth lines. For
large separation $(\xi _{2}-\xi _{1}\gg \xi _{0})$ the exponents in (\ref%
{N_isotrop}) tends to $\delta _{\alpha \beta }(2\sqrt{\pi }\xi _{0}/6)\delta
(\xi _{1}-\xi _{2}),$ and correlation between tangent vectors weakens,\ $%
\left\langle \mathbf{s}^{\prime }(\xi ,t)\mathbf{s}^{\prime }(\xi
,t)\right\rangle \rightarrow 0$, which implies the random walk behavior. \
Thus the characteristic functional with function $\ N$ satisfies to
necessary \textquotedblright semifractal\textquotedblright\ behavior of
closed line \ and will be used further for evaluation of the rate
reconnection coefficients.

Resuming, we exposed main ideas and relations of the gaussian model, which
will be used to calculate intensities of fusion and breakdown of vortex
loop. It will be done in the Appendix B.

\renewcommand{\theequation}{B.\arabic{equation}}
\setcounter{equation}{0} \appendix

\section{APPENDIX B: EVALUATION OF $A(L_{1},L_{2},L)$ AND $B(L,L_{1},L_{2})$}

\subsection{Evaluation of $\ B(l,l_{1},l_{2})$}

We start with the self-intersection processes. Our goal to evaluate $%
B(l,l_{1},l_{2})$ in accordance with relation (\ref{B_def}). Positions of
the line elements $\mathbf{s}(\xi _{2},t),\mathbf{s}(\xi _{1},t)$ and the
relative vector $\mathbf{S}_{b}(\xi _{2},\xi _{1},t)$ are the strongly
fluctuating quantities having the Gaussian statistics. Due to the Wick
theorem \ the average in integrand of (\ref{B_def}) \ can be taken as a sum
of all possible pairs of quantity $\mathbf{S}_{b}(\xi _{2},\xi _{1},t)\,\ \ $%
and its derivatives. Because of uniformity in $\xi $ space, quantity $%
\mathbf{S}_{b}(\xi _{2},\xi _{1},t)\,$\ depends on $\left\vert \xi _{2}-\xi
_{1}\right\vert ,$ for this reason all averages of structure $\left\langle
(\partial X/\partial \xi _{1})\delta (X(\xi _{2},\xi _{1},t))\right\rangle $
vanish, therefore only the pairs separately from $\mathbf{S}_{b}(\xi
_{2},\xi _{1},t)\,$ and from its derivatives survive. As a result the
average of production is equal to production of \ averages and each of the
factors can be evaluated separately
\begin{equation}
\left\langle \left\vert \frac{\partial (X,Y,Z)}{\partial (\xi _{2},\xi
_{1},t)}\right\vert _{\mathbf{\zeta }=\mathbf{\zeta }_{a}}\delta (\mathbf{S}%
_{b}(\xi _{2},\xi _{1},t))\right\rangle =\,\left\langle \left\vert \frac{%
\partial (X,Y,Z)}{\partial (\xi _{2},\xi _{1},t)}\right\vert _{\mathbf{\zeta
}=\mathbf{\zeta }_{a}}\right\rangle \left\langle \delta (\mathbf{S}_{b}(\xi
_{2},\xi _{1},t))\right\rangle .  \label{self_separate}
\end{equation}%
As mentioned, the use of the characteristics functional (\ref{CF_def}), (\ref%
{CF_N}) allows to calculate any averaged functional of configurations $%
\left\{ \mathbf{s}(\xi ,t)\right\} .$ Let us show how to evaluate $%
\left\langle \delta \mathbf{S}_{s}(\xi _{2},\xi _{1},t)\right\rangle $. With
use of the standard integral representation for $\delta $-function
\begin{equation*}
\delta (x)=\frac{1}{(2\pi )}\int_{-\infty }^{\infty }e^{ixy}dy,
\end{equation*}%
we rewrite $\left\langle \delta (\mathbf{S}_{b}(\xi _{2},\xi
_{1},t))\right\rangle $ as
\begin{eqnarray}
\left\langle \delta (\mathbf{S}_{b}(\xi _{2},\xi _{1},t))\right\rangle  &=&%
\frac{1}{(2\pi )^{3}}\int \left\langle \exp \left[ i\mathbf{y(s(}\xi _{2},t%
\mathbf{)-s(}\xi _{1},t\mathbf{))}\right] \right\rangle d^{3}\mathbf{y=}
\notag \\
&&\frac{1}{(2\pi )^{3}}\int \left\langle \exp \left( i\int\limits_{\xi
_{1}}^{\xi _{2}}\mathbf{ys}^{\prime }(\xi ,t)d\xi \ \right) \right\rangle
d^{3}\mathbf{y.}  \label{Delta_integral}
\end{eqnarray}%
Comparing (\ref{CF_def}) and (\ref{Delta_integral}) we conclude that the
integrand in the last term of (\ref{Delta_integral}) is just the
characteristic functional $W(\{\mathbf{P}(\xi ,t)\}),$ taken at value of $%
\mathbf{P}(\xi ,t)$
\begin{equation}
\mathbf{P}(\xi )\;=\;-\mathbf{y}\theta (\xi -\xi _{1})\theta (\xi _{2}-\xi ).
\label{TETA}
\end{equation}%
Here $\theta (\xi )$ is the unit step-wise function. Relation (\ref{TETA})
implies that we choose in integrand of the characteristic functional only
points lying in interval from $\xi _{1}$ to $\xi _{2}$ on \ the curve.
Calculation the average of $\delta (\mathbf{S}(\xi _{2},\xi _{1},t))$ we can
use the model of pure random walk with elementary step equal to $\xi _{0}.$
Practically we can change function \ $N^{\alpha \beta }(\xi _{1}-\xi _{2})$
\ by $\delta _{\alpha \beta }(2\sqrt{\pi }\xi _{0}/6)\delta (\xi _{1}-\xi
_{2})$. According calculation lead to the following result

\begin{eqnarray*}
\left\langle \delta (\mathbf{S}_{b}(\xi _{2},\xi _{1},t))\right\rangle &=&%
\frac{1}{(2\pi )^{3}}\int \exp \left[ -\mathbf{y}^{2}\frac{2\sqrt{\pi }\xi
_{0}}{6}(\xi _{2}-\xi _{1})\right] d^{3}\mathbf{y} \\
&\mathbf{=}&\left( \frac{3\sqrt{3}}{8\pi ^{9/4}}\right) \left( \frac{1}{\xi
_{0}(\xi _{2}-\xi _{1})}\right) ^{3/2}.
\end{eqnarray*}

Evaluation of absolute value of Jacobian in (\ref{self_separate}) we perform
by use of relation $\left\vert J\right\vert =\sqrt{J^{2}}$(See \cite{Rivers})%
$.$ The Jacobian \ consists of production of derivatives with respect to
time and the label variable $\xi $. The latter can be calculated directly
from (\ref{N_isotrop}). Averages including $\mathbf{V}_{l}=d\mathbf{s}/dt$
also can be calculated in explicit form, expressing velocity via the vortex
filament configuration $\{\mathbf{s}(\xi )\}$. However it would be
convenient for the sake of generalization to use the velocity factor $%
\mathbf{V}_{l}=d\mathbf{s}/dt,$ (see comments at the end of this Section).
Calculation of $\ J^{2}$ can be fulfilled writing \ Jacobian in explicit
form and subsequent applying of  the Wick theorem. Simple but tedious
calculations lead to result that
\begin{equation}
J^{2}=\left\langle \mathbf{V}_{lx}^{2}\right\rangle \left\langle (\partial
\mathbf{s}_{y}\mathbf{/}\partial \xi _{1})^{2}\right\rangle \left\langle
(\partial \mathbf{s}_{z}\mathbf{/}\partial \xi _{2})^{2}\right\rangle
+\left\langle \mathbf{V}_{lx}^{2}\right\rangle \left\langle (\partial
\mathbf{s}_{y}\mathbf{/}\partial \xi _{2})^{2}\right\rangle \left\langle
(\partial \mathbf{s}_{z}\mathbf{/}\partial \xi _{1})^{2}\right\rangle +p.p.,
\label{J_squared}
\end{equation}%
where $p.p.$ all permutations with respect to $x,y,z$. \ Taking $%
\left\langle (\partial \mathbf{s}_{y}\mathbf{/}\partial \xi
_{1})^{2}\right\rangle $ and similar terms to be equal to $1/3,$ we obtain
that $\left\vert J\right\vert =\sqrt{12/27}\left\vert \mathbf{V}%
_{l}\right\vert .$ After use of the integration $\int \int d\xi _{1}d\xi
_{2}\delta (\xi _{2}-\xi _{1}-l_{1})\,$\ (see \ref{B_def}) we finally obtain
\begin{equation}
B(l,l_{1},l-l_{1})=b_{s}V_{l}\frac{l}{(\xi _{0}l_{1})^{3/2}},
\label{B_final}
\end{equation}%
where constant $b_{s}=\sqrt{3/64}\pi ^{-9/4}\approx \allowbreak
1.\,\allowbreak 647\,7\times 10^{-2}.$ We introduced in the coefficient $B$
the additional factor $1/2$ to avoid the over-counting of the reconnection
events, since decays $l\rightarrow l_{1}+l_{2}$ and $l\rightarrow l_{2}+l_{1}
$ describe the same process, though the both enter into equations.

\subsection{Evaluation of $A(l_{1},l_{2},l)$}

Let us now evaluate quantity $A(l_{1},l_{2},l)$ defined by relation (\ref%
{A_def}). We again (as for the previous case) evaluate the averages from
Jacobian and $\delta $-function separately. Contribution from Jacobian
coincides with the previous result\ $\left\vert J\right\vert =\sqrt{2}%
\left\vert \mathbf{V}_{l}\right\vert /3$. The rest $\delta $-function part
can be evaluated with the help of the CF obtained above. Unlike the previous
case we have to know two-loop distribution function. Since we omit
interaction of loops (until the reconnection event occurs) the CF \ for two
loops with lengths $l_{1}$and $\ l_{2}$ is just the production of the
expressions of type (\ref{CF_N})

\begin{eqnarray}
W(\{\mathbf{P}_{1}(\xi )\},\{\mathbf{P}_{2}(\xi )\})\; &=&\exp \left\{
-\int\limits_{0}^{l_{1}}\int\limits_{0}^{l_{1}}\mathbf{P}^{\alpha }(\xi
_{1})N_{1}^{\alpha \beta }(\xi _{1}-\xi _{2})\mathbf{P}^{\beta }(\xi
_{2})d\xi _{1}d\xi _{2}\right\} \times   \label{W_double} \\
&&\exp \left\{ -\int\limits_{0}^{l_{2}}\int\limits_{0}^{l_{2}}\mathbf{P}%
^{\alpha }(\xi _{1})N_{2}^{\alpha \beta }(\xi _{1}-\xi _{2})\mathbf{P}%
^{\beta }(\xi _{2})d\xi _{1}d\xi _{2}\right\} .  \notag
\end{eqnarray}%
Quantities $N_{1}^{\alpha \beta }(\xi _{1}-\xi _{2})$ and $N_{2}^{\alpha
\beta }(\xi _{1}-\xi _{2})$ differ from each other only by lengths of loops $%
l_{1}$and $\ l_{2}$, entering expressions for $N_{1}^{\alpha \beta }$.
Further, by use of the standard integral representation for $\delta $%
-function we have
\begin{equation}
\left\langle \delta (\mathbf{S}_{f}(\xi _{2},\xi _{1},t))\right\rangle =%
\frac{1}{(2\pi )^{3}}\int \left\langle \exp \left[ i\mathbf{y(s}_{2}\mathbf{(%
}\xi _{2},t\mathbf{)-s}_{1}\mathbf{(}\xi _{1},t\mathbf{))}\right]
\right\rangle d^{3}\mathbf{y.}  \label{delta_fusion}
\end{equation}%
We stress again that the label variables $\xi _{2}$ and $\xi _{1}$ belongs
two different loops. Let us introduce initial points $\mathbf{s}_{1}\mathbf{(%
}0\mathbf{)}$ and $\mathbf{s}_{2}\mathbf{(}0\mathbf{)}$ and rewrite (\ref%
{delta_fusion}) in the following form:
\begin{equation*}
\frac{1}{(2\pi )^{3}}\int \exp \left[ -i\mathbf{y(s}_{2}\mathbf{(}0\mathbf{%
)-s}_{1}\mathbf{(}0\mathbf{))}\right] \left\langle \exp \left[ i\mathbf{y(s}%
_{2}\mathbf{(}\xi _{2}\mathbf{)-s}_{2}\mathbf{(}0\mathbf{))}\right] \exp %
\left[ -i\mathbf{y(s}_{1}\mathbf{(}\xi _{2}\mathbf{)-s}_{1}\mathbf{(}0%
\mathbf{))}\right] \right\rangle d^{3}\mathbf{y.}
\end{equation*}%
Identifying further \ the \textquotedblright initial\textquotedblright\
positions $\mathbf{s}_{1}\mathbf{(}0\mathbf{),s}_{2}\mathbf{(}0\mathbf{)}$
with quantities $\mathbf{R}_{1},\mathbf{R}_{2}$ in formula (\ref{A_def}) we
rewrite it as
\begin{eqnarray}
A(l_{1},l_{2},l) &=&\frac{1}{\mathcal{V}}\frac{\sqrt{2}}{3}\left\vert
\mathbf{V}_{l}\right\vert \int \int d\mathbf{R}_{1}d\mathbf{R}_{2}\int \int
d\xi _{1}d\xi _{2}  \label{A_inter} \\
&&\frac{1}{(2\pi )^{3}}\int \exp \left[ -i\mathbf{y(\mathbf{R}_{2}-\mathbf{R}%
_{1})}\right] \left\langle \exp \left[ i\mathbf{y(s}_{2}\mathbf{(}\xi _{2}%
\mathbf{)-s}_{2}\mathbf{(}0\mathbf{))}\right] \exp \left[ -i\mathbf{y(s}_{1}%
\mathbf{(}\xi _{2}\mathbf{)-s}_{1}\mathbf{(}0\mathbf{))}\right]
\right\rangle d^{3}\mathbf{y}.  \notag
\end{eqnarray}%
Let us introduce variables $\mathbf{R}_{1}-\mathbf{R}_{2},(\mathbf{R}_{1}+%
\mathbf{R}_{2})/2.$ \ Integration over $\mathbf{\mathbf{R}}_{2}\mathbf{-%
\mathbf{R}}_{1}$ gives $\mathbf{\delta }\mathbf{(y)}$, integration over $(%
\mathbf{R}_{1}+\mathbf{R}_{2})/2$ gives the total volume of system. Further,
integration over $\mathbf{y}$ gives unity, and integration over $\xi
_{1},\xi _{2}$ gives the production $l_{1}l_{2}$. \ Thus we obtain the
remarkable result, that for noninteracting loops the rate coefficient $%
A(l_{1},l_{2},l)$ \ responsible for merging of loops does not depend on
statistics of the individual loop at all and is equal to
\begin{equation}
A(l_{1},l_{2},l)=b_{m}V_{l}l_{1}l_{2}.  \label{A_final}
\end{equation}%
Here $b_{m}=1/\sqrt{18}\approx 0.2357.$ As earlier we introduced additional
factor $1/2$ to avoid the over-counting of the reconnection events.

Results (\ref{B_final}) and (\ref{A_final}) \ (with not well determined
factors $b_{s}$ and $b_{m}$ ) were also obtained in papers \cite{Copeland98}%
, \cite{Steer99}. Authors used some qualitative picture of moving and
colliding elements of lines. This fact confirms the validity of approach
made in our work and allows us to use it for more complicated (in comparison
with the Brownian loops) cases.

Let us discuss the velocity factor $V_{l}$ introduced in relations (\ref%
{B_final},\ref{A_final}). In accordance with formulas (\ref{B_def}), (\ref%
{A_def}), calculation of the coefficients $A(l_{1},l_{2},l)$ and $%
B(l,l_{1},l_{2})$ includes calculations of the time derivatives for the line
elements (velocity factors) and derivatives with respect to the label
variable $\xi $ (structure factors). Various systems such as polymer chains,
cosmic strings, vortex loops have a similar structure, namely, the random
walk or semi-random walk, therefore the structure factors for these systems
are evaluated in a similar manner. The situation with velocity factor is
rather different, which reflects essentially different dynamics for the
different systems. Even for quantized vortices, studied in the present
paper, the velocity of elements can be determined differently depending on
whether the self-induced motion or an external flow prevails in dynamics of
the line (cf. with \cite{Barenghi2004}). In view of the said above and for
the sake of generality, we prefer to use the general velocity factor\ $V_{l}$
instead of calculation in an explicit form. It allows us to extend our
formalism for other systems. Nevertheless we could avoid introductions of
the velocity factor\ $V_{l}$ \ and act as follows. First we have to express
the velocity of the line elements $d\mathbf{s}/dt$ via functional depending
on configuration $\left\{ \mathbf{s}(\xi ,t)\right\} $. This should be done
be use of the motion equation (see e.g., \cite{Donbook}) with the full
Biot-Savart law for the self-induced motion. Then we have to substitute the
components of vector $d\mathbf{s}/dt$ into determinant entering (\ref{B_def}%
), (\ref{A_def}), and calculate the averaged of absolute value of this
determinant with the use of probability functional (\ref{Gauss_model}). \ An
extremely long chain of calculations brings the result that the velocity
factor $V_{l}$ can be estimated as $\ V_{l}$ $=C_{v}$ $\kappa /\xi _{0}$ ,
where $\kappa $ is the quantum of circulation and $C_{v1}$ is a constant
about unity. There is a much simpler (but cruder) way to estimate constant $%
C_{v}$ . Velocity of line element with curvature $\xi _{0}$ in local
approximation and neglecting both the mutual friction and the external flow
is
\begin{equation*}
\mathbf{v}_{l}=\frac{\kappa }{\xi _{0}}\mathbf{n.}
\end{equation*}%
Here $\mathbf{n}$ is the unit vector directed along the binormal \cite{Log}.
The relative velocity of elements in points $\xi _{1\text{ }}$and $\xi _{2%
\text{ }}$is%
\begin{equation}
\mathbf{V}_{l}=(\frac{\kappa }{\xi _{0}}\mathbf{n(\xi }_{1}\mathbf{)-}\frac{%
\kappa }{\xi _{0}}\mathbf{n}(\xi _{2\text{ }})\mathbf{)}  \label{V_rel}
\end{equation}%
In accordance with (\ref{J_squared}) we have to square and to average (\ref%
{V_rel}). Performing it and taking $\left\langle \mathbf{n\mathbf{(\xi }_{1}%
\mathbf{)}n}(\xi _{2\text{ }})\right\rangle =0\,\ $we obtain that velocity
factor $\mathbf{V}_{l}$ is equal $\sqrt{2}\kappa /\xi _{0}$, so constant $%
C_{v}$ for real loop is close to $\sqrt{2}.$

Results of Appendixes A and B are that we calculate $A(l_{1},l_{2},l)$ and $%
B(l,l_{1},l_{2})$ on the base of Gaussian model. Relations (\ref{B_final})
and (\ref{A_final}) are the key results of these Sections. They make the
general "rate equation" (\ref{kinetic equation}) to be the closed problem.


\begin{thebibliography}{99}
\bibitem{uniform} In present work we resrict ourselves by the case, when
network of vortex lines consist of only closed loops. We do not consider
very long lines stretcheng from wall to wall. Thus we deal with the
so-called "uniform" superfluid turbulence. We also do not consider turbulnce
in rotating containers.
\bibitem{loop_1cm} As it is seen from this paper thare are about $20000$
loops of length of $1$ mm and higher. They accumulate about $70\%$ of total
length. The $3D$ size of such loop is of the order $0.3$ mm.
\bibitem{triple} We omitt the rare events when two loops cross each other in
two (or more) different points at the same time, or the single loop
undergoes two (or more) simultaneous self-intersection. These processes may
be important only for the very dense tangle, but the estimations for the
vortex line density requires information about details (e.g., duration) of
the reconnection process.
\bibitem{length-energy} In the local induced approximation the energy of
line is proportional to its length, therefore, instead of the "length"
consevation, we can equally to say about conservation of the energy $E(l,t)$
accumulated in loops of size $l$, instead of the flux of "length", to say
about the flux of energy and so on.
\bibitem{anisotropy} The presence of counterflow velocity violates an
assumptions of the isotrpic Wiener distribution used in previous Sections.
This, however, concerns only relative velocity $\mathbf{V}_{l}$, which is
determined only in order of magnitude (see, Appendix B)
\bibitem{Log} In fact in theis there should present factor of the order $%
(1/4\pi )\ln (\xi _{0}/a_{0})$ ($a_{0}$- is the core size of vortex
filament). In reality this factor is very close to unity and we omit it.
\end{thebibliography}
\end{document}